\documentclass[11pt,a4paper]{article}

\usepackage[pdftex]{graphics}
\usepackage{jheppub}
\usepackage{bm}
\usepackage{float}
\usepackage{amsmath,amssymb,amsfonts}
\usepackage{url}
\usepackage{caption} 
\usepackage{subcaption}
\newcommand{\eq}{\begin{equation}}
\newcommand{\eqe}{\end{equation}}
\newcommand{\eqa}{\begin{eqnarray}}
\newcommand{\eqae}{\end{eqnarray}}

\newcommand{\m}{\Delta}
\newcommand{\M}{\Delta_{\textrm{gap}}}

\newcommand{\LL}{\Lambda_\textrm{max}}

\newbox\charbox
\newbox\slabox

\title{The Geometry of the Modular Bootstrap}

\author{
Li-Yuan Chiang$^{a,b,f}$, Tzu-Chen Huang$^{c}$, Yu-tin Huang$^{a,b}$, Wei Li$^{d}$, Laurentiu Rodina$^{e,h}$, He-Chen Weng$^{a,b,g}$}

\affiliation[a]{Department of Physics and Center for Theoretical Physics, National Taiwan University, Taipei 10617, Taiwan}
\affiliation[b]{Physics Division, National Center for Theoretical Sciences, Taipei 10617, Taiwan}
\affiliation[c]{Enrico Fermi Institute and Kadanoff Center for Theoretical Physics, University of Chicago}
\affiliation[d]{Department of Physics, Boston University, 
Boston, MA  02215, USA}
\affiliation[e]{Centre for Theoretical Physics, Department of Physics and Astronomy, Queen Mary University of London, Mile End Road, London E1 4NS, United Kingdom}
\affiliation[f]{Department of Physics, Yale University, New Haven, CT 06520, USA}
\affiliation[g]{Department of Physics, Brown University, Providence, RI 02912, USA}
\affiliation[h]{Beijing Institute of Mathematical Sciences and Applications, Huairou District, Beijing 101408, China}

\emailAdd{yutinyt@gmail.com}
\emailAdd{weili17@bu.edu}
\emailAdd{l.rodina@qmul.ac.uk}
\emailAdd{he-chen\_weng@brown.edu}
\emailAdd{li-yuan.chiang@yale.edu}

\abstract{We explore the geometry behind the modular bootstrap and its image in the space of Taylor coefficients of the torus partition function. In the first part, we identify the geometry as an intersection of planes with the convex hull of moment curves on $R^+{\otimes}\mathbb{Z}$, with boundaries characterized by the total positivity of generalized Hankel matrices. We phrase the Hankel constraints as a semi-definite program, which has several advantages, such as constant computation time with increasing central charge. We derive bounds on the gap, twist-gap, and the space of Taylor coefficients themselves. We find that if the gap is above $\Delta^*_{\text{gap}}$, where $\frac{c{-}1}{12}<\Delta^*_{\text{gap}}< \frac{c}{12}$, all coefficients become bounded on both sides and kinks develop in the space. In the second part, we propose an analytic method of imposing the integrality condition for the degeneracy number in the spinless bootstrap, which leads to a non-convex geometry. We find that even at very low derivative order this condition rules out regions otherwise allowed by bootstraps at high derivative order.}

\begin{document}

\maketitle

%%%%%%%%%%%%%%%%%%%%%%%%%%%%%%%%%%%%%%%%%%%
\section{Introduction}
For theories whose physical observables can be constrained uniquely via systematic imposition of unitarity, locality and symmetries, one often finds that the observable can be identified with a well defined mathematical entitie. The most well studied example is the amplituhedron for the scattering amplitude of maximal super Yang-Mills~\cite{Arkani-Hamed2014-wv}, where the amplitude is identified as the canonical form on a positive geometry. Other examples include the associahedron for bi-adjoint $\phi^3$ theory~\cite{Arkani-Hamed2018-nd}, the amplituhedron for ABJM theory~\cite{He:2022cup, He:2023rou},  the cosmological polytope for the wave function of conformally coupled scalars~\cite{Arkani-Hamed2017-qj}. 

In some cases general physical principles are not sufficient to uniquely determine specific theories, but instead allow a landscape of possibilities. Remarkably, even this  ``theory space" is in many cases an interesting geometry. A natural question is then how much of allowed space is saturated by known theories, and if the remaining space either contains new unknown theories, or implies some new principle must shrink the space further. 

One such example is the space of Wilson coefficients in the low energy effective field theory of a consistent UV completion. After the pioneering result in \cite{Adams:2006sv}, it was recently uncovered that all coefficients satisfy an infinite number of inequalities, forming the so called EFThedron ~\cite{Arkani-Hamed2021-pa}.  Bounds on these coefficients were obtained in various ways, using both numerical and analytical methods \cite{deRham:2017avq,Caron-Huot:2020cmc,Tolley:2020gtv,Bellazzini:2020cot,Haldar:2021rri,Chiang:2021ziz}. Work on the EFT bootstrap benefited from renewed interest in CFT bootstrap initiated in~\cite{Rattazzi2008-je}, which was able to constrain the conformal dimensions of low lying operators through the method of optimal functionals.\footnote{See also \cite{Bissi:2022mrs,Poland:2022qrs,Hartman:2022zik} for recent reviews on the CFT bootstrap.} The geometric nature of the CFT bootstrap was also explored in \cite{Arkani-Hamed2018-dj,Huang2019-yq}, finding many similarities with the EFThedron. 

In this paper we import recent lessons of the EFT bootstrap and apply them to obtain a geometric interpretation of the modular bootstrap for 2D CFTs ~\cite{Hartman:2014oaa,Benjamin:2016fhe,Cardy:2017qhl,Cho:2017fzo,Dyer:2017rul,Collier:2017shs,Anous:2018hjh,Mukhametzhanov:2019pzy,Benjamin:2019stq,Mukhametzhanov:2020swe,Pal:2020wwd,Benjamin:2020zbs,Dymarsky:2020bps,Lin:2021udi,Benjamin:2021ygh,Grigoletto:2021zyv,Benjamin:2022pnx}. This bootstrap applies the constraint that the CFT can be consistently defined on arbitrary Riemannian manifolds, and so its partition function must satisfy modular invariance of the torus. It was introduced by Hellerman to obtain bounds on the lowest primary~\cite{Hellerman2011-su}, yielding $\frac{c}{6}$ for $c\gg1$, where $c$ is the central charge. This bound was subsequently improved through a series of works~\cite{Friedan:2013cba,Collier:2016cls,Hartman2019-dl}, with the strongest current result $\frac{c}{9.1}$, obtained in \cite{Afkhami-Jeddi2019-ti}. These bounds are weaker than expected from holography where the Banados-Teitelboim-Zanelli (BTZ) black holes~\cite{BTZ} serve as the lowest state of the vacuum theory, having $\Delta=\frac{c}{12}$. While it is now generally believed that the modular bootstrap is not sufficient to improve the bound much further beyond $\frac{c}{9.1}$, several open questions remain, such as the exact implications of integer degeneracy, which has not been systematically studied except in rational CFT~\cite{Kaidi:2020ecu}.

 One perhaps surprising aspect we will quickly observe is that the modular CFT and EFT bootstraps are in fact almost identical mathematical problems. Both bootstraps can be solved efficiently via SDPB, as was done in \cite{Caron-Huot:2020cmc}, but also by phrasing them as moment problems \cite{Moment}. 

To see just how closely related the EFT and CFT bootstraps are, we will show the Taylor coefficients of the partition function $Z(\tau,\bar{\tau})$, expanded around the self dual point $\tau=i$, can be put in a form very similar to the EFT dispersion relations derived in  \cite{Chiang:2021ziz}
\eq\label{CFTeq}
Z_{p,q}^{\textrm{CFT}}=\sum_i n_i \Delta_i^p J_i^q \qquad \qquad g_{p,q}^{\textrm{EFT}}=\sum_i p_i m_i^p J_i^q\,.
\eqe
On the CFT side, $\Delta_i=h_i+\bar{h}_i\ge \M$ is the scaling dimension of state $i$, and $n_i\ge 0$ is the (rescaled) degeneracy of this state. Modular invariance implies that the spin $J_i=h_i-\bar{h}_i$ is an integer, and also that particular linear combinations of $Z_{p,q}$ vanish. On the EFT side, $m_i\ge M$ and $J_i$ are the mass and  spin of UV state $i$, with $M$ being the UV scale, unitarity implies $p_i\ge 0$, while crossing symmetry requires linear combinations of the $g_{k,q}$ to vanish.
The vanishing of these linear combinations are what we will refer to as modular planes, or null planes, and imposing these to high order is one of the main challenges in both bootstrap programs. We will label the highest total derivative order imposed by $\LL\equiv \textrm{max}[p+q]$. Bounds obtained are always valid for any $\LL$, but they become stronger the higher in $\LL$ one can go, until at some point they converge. Because eq.(\ref{CFTeq}) includes spin integrality as a constraint, solving it is referred to as the \emph{spinning} bootstrap. A much simpler problem is ignoring spin altogether, and considering only a subspace of this problem
\eq\label{CFTeqspinless}
Z_{p,0}=\sum_i n_i \Delta_i^p\,,
\eqe
which is referred to as the \emph{spinless} bootstrap. 

Crucially, eq.(\ref{CFTeqspinless}) constitute a single variable moment problem, i.e. what are the sufficient conditions on $Z$
for which the RHS of eq.(\ref{CFTeqspinless}) exists. The solution to such single moment problem is given in terms of positivity conditions on Hankel matrices. Similarly, the expressions in eq.(\ref{CFTeq}) constitute  double moment problems,  with solutions in terms of generalized Hankel matrices. The moment problem formulation provides valid and analytic answers to the bootstrap. At low order in $\LL$, the conditions can even be solved in closed form, but are typically not optimal. However, such solutions may reveal fine details that numerical methods potentially miss. Besides investigating low order closed form results, in this paper we also take a step further in the moment problem approach, and develop a new method to solve the (generalized) Hankel matrix positivity constraints to high order, by phrasing them as a semi-definite program (SDP). This allows us to compare the moment problem solutions of the bootstrap to traditional SDPB approaches at high derivative order. Since  positivity conditions of the (generalized) Hankel matrices are necessary conditions, the space or spectrum that are ruled out attains a \textit{certificate of in-feasibility}, irrespective of the order of derivative- or spin-truncation. 

As an example, we test this setup by computing the gap and comparing to previous results. Our approach allows us to directly compute bounds at arbitrary $c$, for both spinless and spinning bootstrap. In the current paper we have computed up to $c=1200$ and $\LL=11$ for spinning and $c=2000$ and $\LL=43$ for spinless. Evaluating at $c\gg1$ does not introduce any slow down in computation time, which demonstrates the utility of our approach for large $c$ bootstrap.  Much higher $\LL$ can be reached by implementing our program on a cluster.

For spinless, the gap we find is considerably weaker compared to~\cite{Afkhami-Jeddi2019-ti}, which computed up to $\LL=2000$. However, our method allows directly checking much higher $c$, reducing errors due to extrapolation. For the spinning bootstrap, the state of the art was given in~\cite{Collier:2016cls}, at $c=20$ and $\LL=175$. Our results support the findings in~\cite{Collier:2016cls} that imposing spin integrality does not substantially improve bounds on the gap at large central charge.

Next we study the space of consistent CFT, defined in terms of the space of allowed $Z_{p,q}$. In general, this space is unbounded. However, if the spectrum has a gap that is above some threshold, we find the space transitions from unbounded to bounded. This motivates us to term the threshold the critical gap $\Delta^*_{\text{gap}}$. As an example at $\Lambda=11$, the upper bound for the partition function evaluated at the self-dual point $\hat{Z}_0$ is shown below in Figure \ref{intro1} for $c=100$. 
\begin{figure}[H]
    \centering
  \includegraphics[width=1\textwidth]{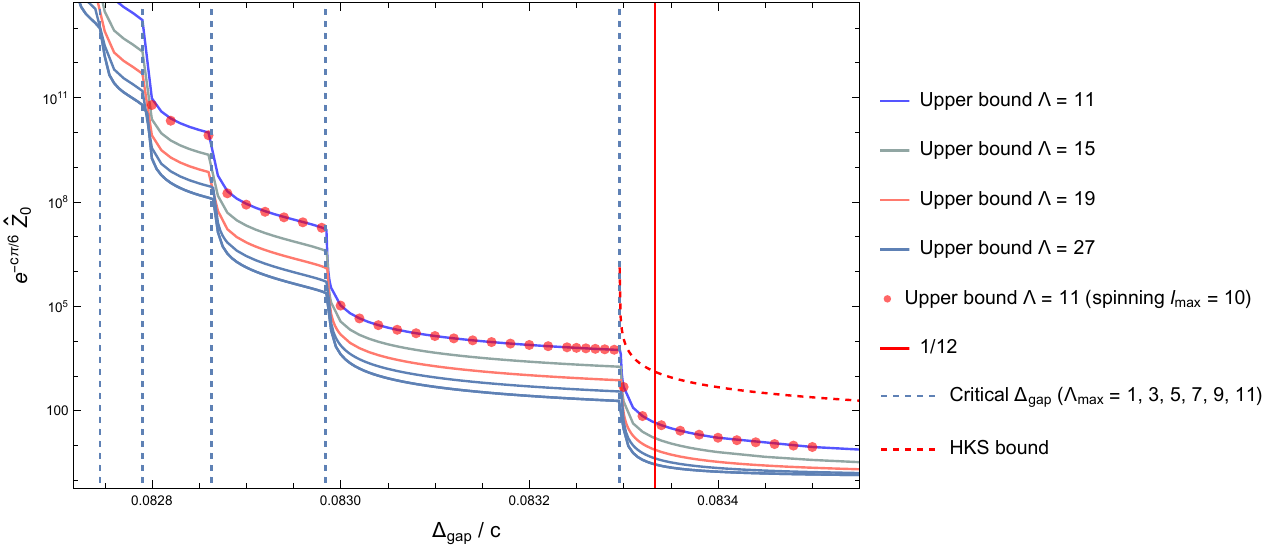}
  \caption{The upper bound on $\hat{Z}_{0}$ plotted against the gap for $c = 100$. There are multiple kinks, situated close to the lower-derivative order critical gaps.}
    \label{intro1}
\end{figure}
We also notice the appearance of kinks between $\Delta^*_{\text{gap}}$ and the maximal gap. These kinks become ubiquitous as $c\gg 1$ , and their position appears stable with respect to derivative truncation, suggesting correlation with physical theories.

In the last part of the paper we take the first steps towards imposing the integer degeneracy condition. While on general grounds it is believed this condition will not affect the gap bounds  at large central charge~\cite{Collier:2016cls}, we will show that it drastically reduces allowed space at $c\sim 1$, even when considered at very low derivative order. We will impose integrality by viewing the partition function as a Minkowski sum of individual states, similar to   the method proposed by some of the authors to impose the unitarity upper bound in the EFT bootstrap \cite{Chiang:2022ltp}. Focusing on the spinless case given by eq.(\ref{CFTeqspinless}), the ``allowed space" of each state $i$ contributing to the partition function is a curve, parameterized by $\Delta_i\in[\M,\infty)$. The allowed space for the whole partition function is therefore a Minkowski sum of such curves, each rescaled by a number $n_i$. We will show that the boundaries of this problem are simple to determine in arbitrary dimension. This more constrained geometry is non-convex, lying inside the previously derived convex geometry, generally touching the original boundary only in isolated points.  

The intersection with modular planes is however more difficult to perform in general. For one null plane, this intersection can be solved analytically, while for more null planes it must be solved numerically. We solve the intersection with two null planes, and find that the integrality condition already excludes extra regions compared to traditional spinless or spinning bootstraps at much higher derivative order. For $c=1$, we find the boundaries significantly approach the points corresponding to the known free boson theories. This is shown in Figure \ref{intro2} below. It is important to note that non-integer bootstraps can only produce convex regions, while at least this $c=1$ case demonstrates that the full picture must necessarily include integrality.
\begin{figure}[H]
    \centering
   \includegraphics[width=1\textwidth]{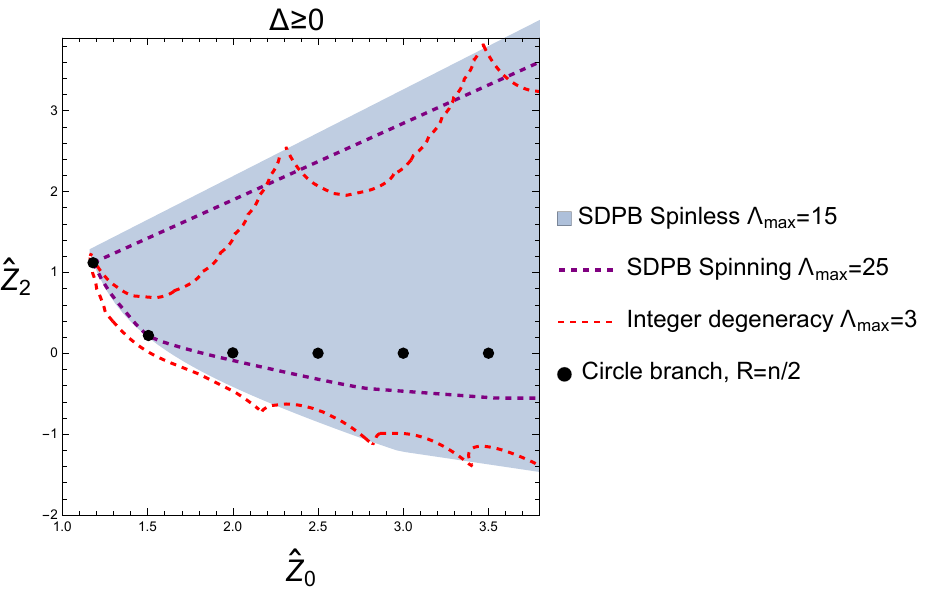}
\caption{Comparing regions carved out by SDPB with the region carved out by the (spinless) degeneracy integrality condition at $\LL=3$. The integrality condition rules out new regions in the space, and  significantly approaches the circle branch theories.}
    \label{intro2}
\end{figure}

At large $c$, the relative effect of integrality with just two null planes appears to wane. However, it is generally expected that high $c$ also requires more null planes for convergence, so computing higher order intersections would be required before making a definitive statement on the role of integrality at high $c$.

To summarize our main results:
\begin{itemize}
\item The spinless/spinning modular bootstrap can be formulated as a single/double moment problem, which is solved in terms of  Hankel matrices
\item At low derivative order, the Hankel positivity can be solved in closed form, while at high order via SDP. This provides some advantages over SDPB, such as eliminating the spin truncation convergence check, and stable efficiency with increasing $c$. 
\item A critical gap exists, above which the allowed space becomes bounded from all directions. Above this gap the space exhibits numerous kinks
\item The integer degeneracy condition leads to a non-convex geometry that significantly reduces allowed space even at very low derivative order. 
\end{itemize}

Added note: During the preparation of this draft, we became aware of the work~\cite{Fitzpatrick:2023lvh} where integrality was imposed over states within a small range of conformal dimensions around the gap. One then requires that at least one integer exists between the lower and upper bounds of the spectral density. It was found that this condition slightly improves the gap at small central charge. In our work, integrality is imposed across the entire spectrum. 

The paper is organized as follows. In Section \ref{section2} we introduce the torus partition function of 2D CFT, which is our object of study. In Section \ref{section3} we describe how the single and double moment problems can be solved in terms of Hankel matrices, and in Section \ref{section4} we show how the partition function can be related to such moment problems. We propose an SDP that can solve positivity of Hankel matrices, and explore bounds on the gap, twist gap, as well as introduce the notion of a critical gap. In Section \ref{section5} we explore the integrality constraint, which we impose by constructing partition function space via iterative Minkowski sums. We compute the intersection with two modular planes, and compare to  results in the previous sections and known theories at $c=1$. We conclude in Section \ref{section6}.

%%%%%%%%%%%%%%%%%%%%%%%%%%%%%%%%%%%%%%%%%%%
\section{Setup}\label{section2}
%%%%%%%%%%%%%%%%%%%%%%%%%%%%%%%%%%%%%%%%%%%
The torus partition function in 2D CFT with central charge $c>1$ is given by
\eq\label{Blocks}
Z(\tau,\bar{\tau})=\chi_0(q)\bar{\chi}_0(\bar{q})+\sum_{h,\bar{h}}n_{h,\bar{h}}\chi_h(q)\chi_{\bar{h}}(\bar{q}),\quad q=e^{i 2\pi\tau}\,,
\eqe
where $\tau$ the torus modulus, and $n_{h,\bar{h}}$ is the degeneracy of the state with scaling dimensions $\Delta=h{+}\bar{h}$ and spin $J=|h{-}\bar{h}|$, i.e. it is a positive number with $n_{0,\bar{0}}=1$. The functions $\chi_h(q)$ are the Virasoro characters, which sum up the contribution to the partition function from the descendants of each Virasoro primary. With $c>1$, they are given by 
\eq
\chi_0(q)=q^{{-}\frac{(c-1)}{24}}\frac{1-q}{\eta(q)},\quad \chi_h(q)=q^{h-\frac{(c-1)}{24}}\frac{1}{\eta(q)}\quad \forall h>0\,,
\eqe
where $\chi_0(q)$ is the vacuum character and $\eta(q)$ is the Dedekind eta function, satisfying $\eta(-\frac{1}{\tau})=\sqrt{-i\tau}\eta(\tau)$.   For $c=1$, degenerate characters appear, which require subtraction, spoiling the  form of eq.(\ref{Blocks}). We leave this generalization to future work, and in this paper focus only on partition functions of the above form, which generally means $c>1$, but also includes a subset of free boson theories at $c=1$, as will be discussed in detail later.

Modular invariance on the torus simply implies invariance under the action of PSL$(2,\mathbb{Z})$ on $(\tau,\bar\tau)$:
\eq\label{modular}
Z(\tau,\bar{\tau})=Z\left(\frac{a\tau{+}b}{c\tau{+}d},\frac{a\bar\tau{+}b}{c\bar\tau{+}d}\right)\,,\quad \left(\begin{array}{cc}a & b \\c & d\end{array}\right)\in {\rm PSL}(2,\mathbb{Z})\,.
\eqe
The group is generated by $S$ and $T$ transformations
\eq
S: \tau \rightarrow -\frac{1}{\tau},\quad T: \tau \rightarrow  \tau{+}1\,.
\eqe
In terms of the representation in eq.(\ref{Blocks}), $T$ invariance is ensured by the spin $J$ being integer. One can simply focus on $S$ invariance $Z(\tau,\bar{\tau})=Z(-\frac{1}{\tau},-\frac{1}{\bar{\tau}})$. This constraint is usually studied by expanding around the $S$ self-dual point 
\eq\label{Spoint}
\left(\tau=i, \;\bar\tau={-}i\right)\,,
\eqe
implying~\cite{Hellerman2011-su}:
\eq\label{Sdual}
\left(\tau \frac{\partial}{\partial \tau}\right)^m \left(\bar\tau \frac{\partial}{\partial \bar\tau}\right)^n Z(\tau,\bar{\tau})|_{\tau=i}=0\,,\quad\forall\, m{+}n\in odd\,.
\eqe
That is, the Taylor expansion allows one to discretize the constraint: instead of a functional identity one now has a vector (which can be taken to be infinite dimensional) identity. 

In the bootstrap program, one imposes modular invariance on eq.(\ref{Blocks}) and extracts constraints on $n_{h,\bar{h}}$ and gaps in $h,\bar{h}$. Here, we would like to approach from another viewpoint, by considering the ``space" of consistent partition functions, defined as $Z(\tau,\bar{\tau})$, that admits a representation of the form eq.(\ref{Blocks}), and respects modular invariance. We begin by discretizing eq.(\ref{Blocks}) around the self-dual point:
\eq\label{Taylor}
z^{(a,b)}=\sum_{h,\bar{h}}n_{h,\bar{h}}\left(\chi^{(a)}_h\right)\left(\chi^{(b)}_{\bar{h}}\right)\,,
\eqe
where $z^{(a,b)}\equiv (\tau\partial_\tau)^a (\bar\tau\partial_{\bar\tau})^b Z|_{\tau=i,\bar{\tau}={-}i}$, $\chi^{(a)}_h\equiv (\tau\partial_\tau)^a \chi_h|_{\tau=i}$. The space of partition functions is now defined by the allowed range for the set of Taylor coefficients 
\eq\label{DiscreteZ}
[\mathbf{Z}]\equiv\left(\begin{array}{cccc}z^{(0,0)} & z^{(0,1)} & z^{(0,2)} & \cdots \\ z^{(1,0)} &  z^{(1,1)} &  z^{(1,2)} & \cdots \\ z^{(2,0)} & z^{(2,1)} & z^{(2,2)} & \cdots \\ \vdots & \vdots & \vdots & \vdots\end{array}\right)\,.
\eqe
By matching order by order in Taylor coefficients on both side of the equality in eq.(\ref{Taylor}), we have:
\eq\label{Convex}
[\mathbf{Z}]= \sum_{h,\bar{h}}n_{h,\bar{h}}\;\;\vec{\chi}_h\;  (\vec{\chi}_{\bar{h}})^T, \quad\quad\quad  \vec{\chi}_h\equiv    \left(\begin{array}{c} \chi^{(0)}_h \\ \chi^{(1)}_h \\ \chi^{(2)}_h \\ \vdots \\ \chi^{(d)}_h\end{array}\right),\, n_{h,\bar{h}}>0\,,
\eqe
where $d$ is the degree of truncation in the Taylor expansion. Since $n_{h,\bar{h}}>0$, the above simply states that the partition function $[\mathbf{Z}]$ must lie within the convex hull of tensored vectors $(\vec{\chi}_{h}, \vec{\chi}_{\bar{h}})$. The convex hull of vectors is a polytope, and here we essentially have the geometry of tensored polytopes. To understand the space we will need to find the boundaries that ``carve" out the hull. 

%%%%%%%%%%%%%%%%%%%%%%%%%%%%%%%%%%%%%%%%%%%
\section{The convex hull of moments}\label{section3}
%%%%%%%%%%%%%%%%%%%%%%%%%%%%%%%%%%%%%%%%%%%
So far the components for the vectors $ \vec{\chi}_h,  \vec{\chi}_{\bar{h}}$ on the RHS of eq.(\ref{Convex}) are polynomials in $h$. As we will soon see, after a series of simple GL transformations, these can be transformed into points on a $d$-dimensional moment curve, i.e. each vector is a collection of monomials with increasing degree:
\eq
\left(\begin{array}{c}1 \\ h \\h^2\\ \vdots \\ h^d\end{array}\right)\,.
\eqe  
The convex hull of such moments is well discussed in mathematical literature, and is termed the moment problem. In its most general form the moment problem identifies the necessary and sufficient conditions on a vector $\vec{y}=(y_0, y_1,\cdots, y_n)$, such that 
\eq
y_n=\int_I \rho(x) x^n dx,\quad x\in I\,,
\eqe
admits a solution  $\rho(x)$ that is non-negative. Here $I$ is some region on the real axes, which falls into three categories: Hamburger $I\in(-\infty, +\infty)$, Stieltjes $I \in[0,+\infty)$ and Hausdorff $I \in[0,1]$ moment problem. Equivalently, these conditions are then the co-dimension one boundaries for the convex hull of points on a moment curve. For finite $n$, this is called the truncated moment problem, and sufficient conditions can be given phrased in terms of the symmetric Hankel matrix of $y_n$. 

One can generalize to the bi-variate moment problem, which will be relevant for the spinning bootstrap. In this case, we ask for conditions on $[y]_{n\times n}$ such that 
\eq
y_{n,m}=\int_I \rho(x, y) x^n y^m \,dxdy ,\quad x,y\in I\,,
\eqe
admits a non-negative solution for $\rho(x, y)$. For the bi-variate problem, only necessary conditions are known for the most general truncated moment problems, if we require a finite number of constraints. However, we will numerically demonstrate that these conditions lead to constraints that are close to the true boundary.  

%%%%%%%%%%%%%%%%%%%%%%%%%%%%%%%%%%%%%%%%%%%
\subsection{The single moment problem}\label{sec:SingleMom}
%%%%%%%%%%%%%%%%%%%%%%%%%%%%%%%%%%%%%%%%%%%
Let us first review the geometry associated with the convex hull of moment curves, or the single moment problem:
\eq\label{MomentCurveHull}
\vec{y}=\left(\begin{array}{c} y_0 \\ y_1 \\ y_2 \\ \vdots \\ y_d\end{array}\right)=\int_I \rho(x) \left(\begin{array}{c} 1 \\ x \\ x^2 \\ \vdots \\ x^{d}\end{array}\right)dx,\quad \,\rho(x)>0,\quad \forall x\in I\,.
\eqe
As shown in~\cite{Orbitopes}, the image of the hull is carved out by the statement that the symmetric Hankel matrix $K[\vec{y}]$, defined as 
\eq
K_n[\vec{y}]\equiv \left(\begin{array}{cccc} y_0 & y_1 & \cdots & y_{n} \\y_1 & y_2  & \cdots  & y_{n{+}1} \\ \vdots & \vdots & \vdots   & \vdots  \\ y_n & y_{n{+}1} & \cdots & y_{2n}  \end{array}\right)\,,
\eqe
is a positive semi-definite matrix, which implies its leading principle minors are non-negative. If we further constrain ourselves to the convex hull of half moment curves, where $x_i\in R^+$, we then have that the Hankel and shifted Hankel matrix are totally positive matrices, i.e. all minors are non-negative. The independent conditions are given as:
\eq\label{ConvHalf}
{\rm Det}\,\left( K_n[\vec{y}]\right)\geq0, \quad   {\rm Det}\,\left( K^{\rm shift}_n[\vec{y}]\right)\equiv{\rm Det}\,\left( K_n[\vec{y}]\right)|_{y_i\rightarrow y_{i+1}}\geq0\,.
\eqe
That these are necessary conditions can be derived as follows: the Hankel matrices can be written as 
\eq
K_{n}=\int_I\; \rho(x)\,\vec{x}\vec{x}^T \;dx\,,
\eqe
where $\vec{x}^T=(1,x,\cdots,x^d)$. Sandwiching $K_{n}$ with a random vector one easily sees that $\vec{v}^T K_n \vec{v}\geq0$, and hence $K_n$ is a positive semi-definite matrix. The shifted Hankel can be written as:
\eq
K^{\rm shift}_n=\int_I\;  x\rho(x)\,\vec{x}\vec{x}^T \;dx\,.
\eqe
If $x\in \mathbb{R}^+$ then $x\rho(x)\geq0$, and one similarly concludes that $K^{\rm shift}_n$ is a positive semi-definite matrix.

Since the boundaries are associated with singular Hankel matrices, this implies that the boundary is parameterized by the convex hull of finite elements. Consider $n=2p{+}1$, where the space is carved out by 
\eq
K_{m}\succeq 0, \quad K^{\rm shift}_{m}\succeq 0,\quad \forall m\leq p\,.
\eqe
The space is projective $\mathbb{P}^{2p{+}1}$ and the co-dimension one boundary is naturally associated with ${\rm Det}K_p=0$ or ${\rm Det}K^{\rm shift}_p=0$. We can readily read off the spectrum on these boundaries
\eqa\label{eq: BoundaryStates}
{\rm Det}K_p&=&0,\quad (\Delta_1,\cdots,\Delta_p, \Delta_\infty) \,,\nonumber\\
{\rm Det}K^{\rm shift}_p&=&0,\quad (\Delta_{\rm gap}, \Delta_1,\cdots,\Delta_p)\,,   
\eqae
where $\Delta_i$ represents a state with $(1, \Delta_i, \Delta^2_i, \cdots \Delta^{2p{+}1}_i)$, $\Delta_\infty$ being $(0,0,\cdots, 0,1)$ and $\Delta_{\rm gap}$ corresponds to the state $(1,0,0,\cdots,0)$. The set of states in eq.(\ref{eq: BoundaryStates}) can be easily understood from the fact that $\Delta_\infty$ do not contribute to  $K_p$ while $\Delta_{\rm gap}$ do not contribute to $K^{\rm shift}_p$. The free parameters that parameterize the hull are the $2p$ parameters $(\Delta_i,\rho_i)$ and $\rho_\infty$ or $\rho_{\rm gap}$, with one degree of freedom removed due to the projective nature.

%%%%%%%%%%%%%%%%%%%%%%%%%%%%%%%%%%%%%%%%%%%
\subsection{The double moment problem}
%%%%%%%%%%%%%%%%%%%%%%%%%%%%%%%%%%%%%%%%%%%
We now move on to the bi-variate problem. More precisely, we are interested in the space of $\left[\mathbf{Y}\right]_{d{+}1\times d{+}1}$ given by:
\eq\label{Prod}
\left[\mathbf{Y}\right]_{d{+}1\times d{+}1}=\left(\begin{array}{cccc}y^{(0,0)} & y^{(0,1)}  & \cdots & y^{(0,d)}  \\  y^{(1,0)}  & y^{(1,1)}  & \cdots & y^{(1,d)}  \\ \vdots  & \vdots & \vdots & \vdots \\ y^{(d,0)} & y^{(d,1)} & \cdots & y^{(d,d)}\end{array}\right)=\int_{I}\;\rho(x,\tilde{x})\vec{x}\vec{\tilde{x}}^T dx d\tilde{x}\,,
\eqe
where $\rho(x,\tilde{x})\geq 0$ for $x,\tilde{x}\in I$ exists. For each fixed row or column, we can construct a Hankel and shifted-Hankel matrix which are positive semi-definite from our previous argument. Furthermore,  consider the following ``generalized Hankel matrix"
\eq\label{GenHank}
\mathbf{K}[\mathbf{Y}]\equiv \left(\begin{array}{ccccc}y^{(0,0)} & y^{(0,1)} & y^{(1,0)}& y^{(0,2)} &\cdots \\ y^{(0,1)} &   y^{(0,2)} &  y^{(1,1)} & y^{(0,3)} &\cdots \\ y^{(1,0)} & y^{(1,1)} & y^{(2,0)} & y^{(1,2)} &\cdots  \\ y^{(0,2)} & y^{(0,3)} & y^{(1,2)} & y^{(0,4)} &\cdots  \\ \vdots & \vdots & \vdots & \vdots &\cdots\end{array}\right)\,.
\eqe
If $d=2m{+}1$, the generalized Hankel matrix will be $(m{+}1)(2m{+}3){\times}(m{+}1)(2m{+}3)$. If a solution to eq.(\ref{Prod}) exists, we claim that the generalized Hankel is also a positive semi-definite matrix. Indeed substituting the RHS of eq.(\ref{Prod}) into $\mathbf{K}[\mathbf{Y}]$ we see that 
\eq
\mathbf{K}[\mathbf{Y}]=\int_{I}\;\rho(x,\tilde{x}) \vec{\mathbf{X}} (\vec{\mathbf{X}} )^T \;dxd\tilde{x}\,, 
\eqe
where $(\vec{\mathbf{X}} )^T= (1,x, \tilde{x}, x^2, \tilde{x} x, \tilde{x}^2,\cdots)$. Then, since 
\eq
\mathbf{V}^T\mathbf{K}[\mathbf{Y}]\mathbf{V}=\int_{I}\;\rho(x,\tilde{x})  (\mathbf{V}^T \vec{\mathbf{X}})^2 \geq 0\,,
\eqe
for any real vector $\mathbf{V}$, we conclude that  $\mathbf{K}[\mathbf{Y}]$ is a positive semi-definite matrix. If we further restrict that $x, \,\tilde{x}\in \mathbb{R}^+$ implies that the shifted generalized Hankel matrices, either by one column or one row,
\eqa\label{Kshift}
\mathbf{K}[\mathbf{Y}]^{{\rm shift}_q}=\mathbf{K}[\mathbf{Y}]|_{y^{(p,q)}\rightarrow y^{(p,q{+}1)}}, \quad \mathbf{K}[\mathbf{Y}]^{{\rm shift}_p}=\mathbf{K}[\mathbf{Y}]|_{y^{(p{+}1,q)}\rightarrow y^{(p,q)}}\,, 
\eqae
must also be positive semi-definite. For the bi-variate moment problem, the vanishing of the minors for the generalized Hankel no longer leads to simple constraint on the number of elements in the hull. To see this consider 
\begin{equation}\label{eq:Counter}
{\rm Det}\left(\begin{array}{cc}y^{(0,0)} & y^{(0,1)}  \\y^{(0,1)} & y^{(0,2)} \end{array}\right)\,.
\end{equation}
This can vanish with an infinite number of states with distinct $x$ but the same $\tilde{x}$.

The sufficient conditions are less understood for the bi-variant moment problem. On general grounds, we expect that the boundaries are parameterized by finite number of elements in the hull. That is, the degrees of freedom that parameterize the boundary should be the $(\rho_i, x_i, \tilde{x}_i)$s of the finite elements. For the single moment problem, the Hankel matrix being singular puts a bound on the number of elements, and thus they serve as the boundary or define limiting points to the boundary. For generalized Hankel of the double moment, this is no longer true. At truncated order, we no longer expect that the singular generalized Hankel describes the true boundary. However, if we have a collection of  singular Hankel matrices, one can infer a bound on the number of elements. Indeed, this expectation is verified when compared to numerical SDPB results in the next section.

%%%%%%%%%%%%%%%%%%%%%%%%%%%%%%%%%%%%%%%%%%%
\section{The modular-hedron}\label{section4}
%%%%%%%%%%%%%%%%%%%%%%%%%%%%%%%%%%%%%%%%%%%
\label{the_modular-hedron}
We are now ready to map the modular bootstrap problem to the moment problem. It will be more convenient to begin with the reduced partition function $\hat{Z}(\tau,\bar{\tau})$, defined as
\begin{equation}\label{reduceS}
\hat{Z}(\tau,\bar{\tau})=|\tau|^{\frac{1}{2}}|\eta(\tau)|^2Z(\tau,\bar{\tau})\,.
\end{equation}
As the pre-factor $|\tau|^{1/2}|\eta(\tau)|^2$ is invariant under $S$-transform $\tau\to-1/\tau$, the reduced partition function retains its invariance. Similarly, the expansion is now defined on the reduced characters, which are independent of the $\eta$ functions: 
\begin{equation}\label{Scharacter}
\hat{Z}(\tau,\bar{\tau})=\hat{\chi}_0(\tau)\hat{\bar{\chi}}_0(\bar{\tau})+\sum_{h,\bar{h}>0}n_{h,\bar{h}}\hat{\chi}_h(\tau)\hat{\bar{\chi}}_{\bar{h}}(\bar{\tau})\,,
\end{equation}
where
\begin{equation}
\hat{\chi}_0(\tau)\hat{\chi}_0(\bar{\tau})=|\tau|^{1/2}|q^{{-}\frac{c{-}1}{24}}(1-q)|^2,\quad \hat{\chi}_h(\tau)\hat{\chi}_{\bar{h}}(\bar{\tau})=|\tau|^{1/2}q^{h-\frac{c-1}{24}}\bar{q}^{\bar{h}-\frac{c-1}{24}}\,.
\end{equation}
Again we expand both sides of eq.(\ref{Scharacter}) around the self-dual point. First for the reduced characters,
\eqa\label{Polyfunc}
\hat{\chi}^{(n)}_h\equiv\left.(\tau \partial_{\tau})^n \hat{\chi}_h(\tau)\right|_{\tau=i}&=&(-1)^\frac{1}{8}e^{\frac{(c{-}1{-}24 h)\pi}{12}}(a_{n,n}{+}X(\cdots(a_{n,3}{+}X(a_{n,2}{+}X(a_{n,1}{+}X))\,, \nonumber\\
\hat{\chi}^{(n)}_{\bar{h}}\equiv\left.(\bar\tau \partial_{\bar\tau})^n \hat{\chi}_{\bar{h}}(\bar\tau)\right|_{\bar\tau={-}i}&=&-(-1)^\frac{7}{8}e^{\frac{(c{-}1{-}24 \bar{h})\pi}{12}}(a_{n,n}{+}\bar{X}(\cdots(a_{n,3}{+}\bar{X}(a_{n,2}{+}\bar{X}(a_{n,1}{+}\bar{X}))\,,\nonumber\\
\eqae
where $X=(c{-}1{-}24 h)\pi$, $\bar{X}=(c{-}1{-}24\bar{h})\pi$ and $a_{n,q}$ are numbers independent of $(c, h, \bar{h})$. Up to a universal pre-factor $\alpha_{c,h}\equiv({-}1)^\frac{1}{8}e^{\frac{(c{-}1{-}24 h)\pi}{12}}$ and $\alpha_{c,\bar{h}}\equiv{-}({-}1)^\frac{7}{8}e^{\frac{(c{-}1{-}24 \bar{h})\pi}{12}}$, we see that the $\hat{\chi}^{(n)}_h$s ( $\hat{\chi}^{(n)}_{\bar{h}}$) are proportional to a polynomial of degree $n$ in $h$ $(\bar{h})$. As an example
\eqa\label{Components}
\hat{\chi}^{(0)}_h&=&\alpha_{c,h},\quad \hat{\chi}^{(1)}_h=\alpha_{c,h}\frac{1}{12}(3{+}X)\,,\nonumber\\
\hat{\chi}^{(2)}_h&=&\alpha_{c,h}\frac{1}{(12)^2}(9{+}X(18{+}X)),\quad \hat{\chi}^{(3)}_h=\alpha_{c,h}\frac{1}{(12)^3}(27{+}X(279{+}X(45{+}X))\,.
\eqae
Collecting $\hat{\chi}^{(n)}_h$ into a vector, $\vec{\hat{\chi}}_h=(\hat{\chi}^{(0)}_h, \hat{\chi}^{(1)}_h,\cdots)$, via a universal $c$ dependent GL transformation $M(c)$ we can rotate it into a moment form. For example, truncating at the third derivative order :
\eqa
&&M(c)=\left(\begin{array}{ccc}
 1 & 0 & 0 \\
 \frac{\pi  (c-1)+3}{24 \pi } & -\frac{1}{2 \pi } & 0 \\
 \frac{\pi  (c-1) (\pi  (c-1)+6)+45}{576 \pi ^2} & \frac{-\pi  c+\pi -9}{24 \pi ^2} & \frac{1}{4 \pi ^2} \\
\end{array}\right)\nonumber\\
&& \quad\rightarrow \quad M(c) \left(\begin{array}{c} \hat{\chi}^{(0)}_h \\ \hat{\chi}^{(1)}_h \\ \hat{\chi}^{(2)}_h \end{array}\right)=\alpha_{c,h}M(c)\left(\begin{array}{c} 1 \\ \frac{1}{12}(3{+}X) \\ \frac{1}{(12)^2}(9{+}X(18{+}X)) \end{array}\right)= \alpha_{c,h}\left(\begin{array}{c} 1 \\ h \\ h^2 \end{array}\right)\nonumber\,.\\
\eqae
Applying the same double expansion for the partition function and the vacuum character, eq.(\ref{Scharacter}) can be reorganized as: 
\eq\label{eq: ModularHedron}
\framebox[12cm][c]{$\displaystyle [\widetilde{\mathbf{Z}}]_{h, \bar{h}}=M(c)\left([\mathbf{\hat{Z}}]{-}\vec{\hat{\chi}}_0 \vec{\hat{\chi}}^{\,T}_0\right)M^{\,T}(c)=\sum_{h,\bar{h}} \; \tilde{n}_{h,\bar{h}} \,\left(\begin{array}{c} 1 \\ h \\ h^2 \\ \vdots\end{array}\right)\left(\begin{array}{c} 1 \\ \bar{h} \\ \bar{h}^2 \\ \vdots \end{array}\right)^{T}$\,,}
\eqe
where $\tilde{n}_{h,\bar{h}}=n_{h,\bar{h}}e^{\frac{(c{-}1{-}12(h{+}\bar{h}))\pi}{6}}$. The subscript for $ [\widetilde{\mathbf{Z}}]_{h, \bar{h}}$ indicates the double moment is in $(h, \bar{h})$. We see that after the double GL rotation, the vacuum-subtracted partition function is given by the convex hull of a tensor product of half-moment curves. We will term this space the ``\textit{modular-hedron}".  It is sometimes more convenient to consider states labeled by $(\Delta, \mathcal{J})$. The geometry for $[\widetilde{\mathbf{Z}}]_{\Delta,\mathcal{J}}$ is simply a linear transform of $[\widetilde{\mathbf{Z}}]_{h,\bar{h}}$. For example, 
\eqa
&&\tilde{z}_{\Delta,\mathcal{J}}^{(0,0)}=\tilde{z}_{h, \bar{h}}^{(0,0)},\quad \tilde{z}_{\Delta,\mathcal{J}}^{(0,1)}=\tilde{z}_{h, \bar{h}}^{(1,0)}{-}\tilde{z}_{h, \bar{h}}^{(0,1)}, \quad \tilde{z}_{\Delta,\mathcal{J}}^{(1,0)}=\tilde{z}_{h, \bar{h}}^{(1,0)}{+}\tilde{z}_{h, \bar{h}}^{(0,1)}\,.
\eqae
In this basis, the geometry is associated with the tensor product of a half- and a discrete moment curve, where $\mathcal{J}\in \mathbb{Z}$.  \\

\noindent \textbf{The spinless modular polytope}

A simple limit of the geometry is the case where we only consider $\tilde{z}^{(q,0)}$, i.e. where the spin information is suppressed. Then the double moment reduces back to the single moment. This corresponds to the spinless geometry where we consider the torus moduli being purely imaginary, $\tau=i\beta$ and
\eq\label{eqn:spinlessZ}
Z[\beta]=\sum_{\Delta} n_\Delta e^{2\pi\beta\left(\Delta-\frac{c{-}1}{12}\right)}\,.
\eqe
We then have the modular-polytope 
\eq\label{eq: ModularPoly}
\vec{\mathcal{Z}}^{\rm spinless}=\sum_\Delta n_\Delta 
\begin{pmatrix}
1\\
\Delta\\
\vdots\\
\Delta^d
\end{pmatrix}\,,
\eqe
where we identify $\mathcal{Z}^{\rm spinless}_q=\tilde{z}_{\Delta, \mathcal{J}}^{(q,0)}=\sum_{i=0}^q \begin{pmatrix}
q\\i
\end{pmatrix} \tilde{z}_{h, \bar{h}}^{(i,q-i)}$.

The character representation implies that the Taylor coefficients of the partition function must lie \textit{inside} the hull of tensored moment curves as in eq.(\ref{eq: ModularHedron}). This  geometry encodes the positivity of the degeneracy number, as well as unitarity $h, \bar{h}\geq0$. Modular invariance is then reflected in the additional requirement that $\hat{z}^{(i,j)}=0$ for $i{+}j\in odd$. That is, the discretized partition function lives on a ``\textit{modular subplane}" 
\eq
[\mathbf{\hat{Z}}]_{\rm mod}\equiv\left(\begin{array}{ccccc} \hat{z}^{(0,0)} & 0 & \hat{z}^{(0,2)} & 0& \cdots \\ 0 &  \hat{z}^{(1,1)}&  0 & \hat{z}^{(3,1)}&\cdots \\ \hat{z}^{(2,0)} & 0 & \hat{z}^{(2,2)} & 0&\cdots \\ 0 & \hat{z}^{(3,1)}& 0 & \hat{z}^{(3,3)}&\cdots \\ \vdots & \vdots & \vdots & \vdots&\end{array}\right)\,.
\eqe
The geometry for the modular bootstrap is an intersection problem, taken between the modular plane $[\mathbf{\hat{Z}}]_{\rm mod}$ with the modular-hedron $[\mathbf{\hat{Z}}]$. The boundary of this intersection will be determined by the boundaries of the modular-hedron. Restricting ourselves to the spinless geometry, i.e. the modular-polytope, this immediately implies that for $d=2p{+}1$, the boundary of the intersection would be given by $p$ states, as observed in~\cite{Afkhami-Jeddi2019-ti}.

In the following, we will systematically analyze the geometry using both analytical and numerical methods. We will also impose further constraints on the spectrum such as, 
\begin{itemize}
    \item A gap for either $\Delta$ or $min[h,\bar{h}]$.
    \item Spin integrality.
    \item Integrality for the degeneracy number.
\end{itemize}
This will allow us to glean information with regards to the twist-gap, gap, and bounds on the value of the partition function at the self-dual point.

\subsection{The semi-definite program setup}
Here we summarize the constraints developed in the previous sections as follows:
\begin{itemize}
  \item Modular-hedron geometry:
    \begin{equation}\label{constraint_1}
      \mathbf{K}\left[[\widetilde{\mathbf{Z}}]_{h,\bar{h}}\right] \succeq 0\,.
    \end{equation}
    where $[\widetilde{\mathbf{Z}}]_{h,\bar{h}}$ is defined in eq.(\ref{eq: ModularHedron}), and $\mathbf{K}$ is defined in eq.(\ref{GenHank}).
  \item Unitarity: \(h, \bar{h} \geq 0\)
      \begin{equation}\label{constraint_2}
        \mathbf{K}\left[[\widetilde{\mathbf{Z}}]_{h,\bar{h}}\right]^{\text{shift}_{h}} \succeq 0, \quad \mathbf{K}\left[[\widetilde{\mathbf{Z}}]_{h,\bar{h}}\right]^{\text{shift}_{\bar{h}}} \succeq 0\,.
      \end{equation}
  \item Integer spin: \(\mathcal{J} \in \mathbb{Z}\)
      \begin{equation}
        \mathbf{K}\left[[\widetilde{\mathbf{Z}}]_{h,\bar{h}}\right]^{l, l+1} \succeq 0 \quad \forall l \in \mathbb{Z} \,,
      \end{equation}
  where \(\mathbf{K}[\tilde{\mathbf{Z}}]^{l, l+1}\) is given in eq.(\ref{Kspin}).
  \item Modular invariance: 
    \begin{equation}\label{constraint_4}
      \hat{z}^{(i,j)}=0 \quad \forall \, i{+}j \in \text{odd}\,.
    \end{equation}
\end{itemize}
These constraints, together with a linear function of \(\hat{z}^{(i,j)}\)  to be extremized, constitute a semi-definite program (SDP). In particular, we begin with the relevant Hankel matrices parameterized by the modular plane, i.e. $\hat{z}^{(i,j)}$ with $i{+}j=even$, with suitable deformations to incorporate a gap for the spectrum. For fixed values of gap, this sets up a feasibility problem for SDP. In addition to finding the optimal value for the discretized partition function, it is also possible to obtain the optimal gap. The idea is to make an assumption about the gap and then find a set of \(\hat{z}^{(i,j)}\) that satisfies all the constraints. If a certificate of infeasibility to the problem is found, the proposed spectrum is then ruled out. See Appendix \ref{appendix_SDP} for a detailed discussion of SDP and the feasibility problem.

In our notation, the traditional SDPB approach starts with the modular constraint:
\begin{equation}
\sum_{h,\bar{h}} n_{h,\bar{h}}f_{a,b}(h,\bar{h})\equiv \sum_{h,\bar{h}} n_{h,\bar{h}}\chi_h^{a} \chi_{\bar{h}}^{(b)}=0\,\quad \forall\; a{+}b={\rm odd} \,.
\end{equation}
One then constructs linear combinations of $f_{a,b}(h,\bar{h})$ that are positive above a gap. Each such functional rules out any spectrum whose lowest twist (or dimension) lives above the corresponding gap. The combination that gives a minimum gap is then the optimal functional. Translated to our geometry, what the SDBP probes is the intersection geometry of a co-dimension $n$-modular plane, where $n$ is the number of modular constraints, with the \textit{infinite dimensional} double moment problem. By comparing the result to our geometric bounds, it provides a measure of how far the generalized Hankel constraints are from the true boundary.

%%%%%%%%%%%%%%%%%%%%%%%%%%%%%%%%%%%%%%%%
\subsection{Bounding the gap}
%%%%%%%%%%%%%%%%%%%%%%%%%%%%%%%%%%%%%%%%%%
As a warm up, using the semi-definite programming setup in the previous subsection, we study the gap and twist gap at fixed derivative order up to $\Lambda=43$. This is far from $\Lambda\sim 2000$ studied in ~\cite{Afkhami-Jeddi2019-ti}, and we do not expect novel bounds. The main purpose is to demonstrate the feasibility of this approach and comparing spin-less vs spinning results as well as the effects of integrality of spins. 

To impose a gap on the spectrum, recall that total positivity of the Hankel matrices stems from its representation as a convex hull: 
\begin{equation}
K\left[[\widetilde{\mathbf{Z}}]_{h,\bar{h}}\right]= \sum_{h, \bar{h}} \tilde{n}_{h, \bar{h}}
  \begin{pmatrix}
    1 \\ h \\ \bar{h} \\ \vdots
  \end{pmatrix}
  \begin{pmatrix}
    1 \\ h \\ \bar{h} \\ \vdots
  \end{pmatrix}^T\,.
\end{equation}
If the spectrum has a gap, say $\Delta_{\rm gap}$, then we immediately see that 
\begin{equation}\label{eq: HankelDeform}
K\left[[\widetilde{\mathbf{Z}}]_{h,\bar{h}}\right]^{{\rm twist}_\Delta}=\sum_{h, \bar{h}} \tilde{n}_{h, \bar{h}}(h{+}\bar{h}-\Delta_{\rm gap})
  \begin{pmatrix}
    1 \\ h \\ \bar{h} \\ \vdots
  \end{pmatrix}
  \begin{pmatrix}
    1 \\ h \\ \bar{h} \\ \vdots
  \end{pmatrix}^T\,,
\end{equation}
also constitute a definite positive matrix. In terms of the original coordinates, we can for example write:\begin{equation}
\begin{pmatrix}
    \tilde{z}^{(0,1)} & \tilde{z}^{(1,1)} & \tilde{z}^{(0,2)} \\
    \tilde{z}^{(1,1)} & \tilde{z}^{(2,1)} & \tilde{z}^{(1,2)} \\
    \tilde{z}^{(0,2)} & \tilde{z}^{(1,2)} & \tilde{z}^{(0,3)}
  \end{pmatrix}+\begin{pmatrix}
  \tilde{z}^{(1,0)} & \tilde{z}^{(2,0)} & \tilde{z}^{(1,1)} \\
  \tilde{z}^{(2,0)} & \tilde{z}^{(3,0)} & \tilde{z}^{(2,1)} \\
  \tilde{z}^{(1,1)} & \tilde{z}^{(2,1)} & \tilde{z}^{(1,2)}
\end{pmatrix}
- \Delta_{\text{gap}}
\begin{pmatrix}
  \tilde{z}^{(0,0)} & \tilde{z}^{(1,0)} & \tilde{z}^{(0,1)} \\
  \tilde{z}^{(1,0)} & \tilde{z}^{(2,0)} & \tilde{z}^{(1,1)} \\
  \tilde{z}^{(0,1)} & \tilde{z}^{(1,1)} & \tilde{z}^{(0,2)}
\end{pmatrix}
\succeq 0\,.
\end{equation}
For a fixed truncation order $\Lambda_{\text{max}}$, i.e. the total degree in $h$ and $\bar{h}$, we solve the feasibility problem for various different values of the gap. For the spinless bootstrap, $\Lambda_{\text{max}}$ is just the highest degree in $\Delta$. This enables us to determine an upper bound for \(\Delta_{\text{gap}}\), and similarly for the twist-gap. 

It is also straightforward to impose integrality on the spins. We will define $\mathcal{J}\equiv h{-}\bar{h}\in \mathbb{Z}$.  Suppose the spin \(\mathcal{J}\) does not take value between \(l\) and \(l + 1\), the following matrix will be positive:
\begin{equation}\label{Kspin}
  \mathbf{K}\left[[\widetilde{\mathbf{Z}}]_{h, \bar{h}}\right]^{l, l + 1} = \sum_{h, \bar{h}} \tilde{n}_{h, \bar{h}} (h{-}\bar{h}{-}l) (h{-}\bar{h}{-}l {-} 1)
  \begin{pmatrix}
    1 \\ h \\ \bar{h} \\ \vdots
  \end{pmatrix}
  \begin{pmatrix}
    1 \\ h \\ \bar{h} \\ \vdots
  \end{pmatrix}^T \succeq 0\,.
\end{equation}
By imposing the positivity of $\mathbf{K}[\widetilde{\mathbf{Z}}]^{l, l+1}$ for each pair of adjacent spins \((l, l+1)\), one effectively requires that \(\mathcal{J}\) take only integer values. In practice, to implement these constraints, one considers only a finite number of spins $-l_{\text{max}} \leq l \leq l_{\text{max}}$. It is however worth noting that the Hankel matrix positivity, in contrast with the polynomial program approach of SDPB, gives valid upper bounds on $\Delta_{\text{gap}}$ for any spin-truncation $l_{\text{max}}$ and truncation order. Increasing $l_{\text{max}}$ only \emph{improves} the bound. This is because the positivity of the truncated Hankel matrices is always a necessary condition for the double-moment geometry, and we are bounding the gap by ruling out configurations of the spectrum. \\

\noindent \textbf{Bounding the gap of scaling dimension}
%%%%%%%%%%%%%%%%%%%%%%%%%%%%%%%%%%%%%%%%%%%%%%%%%%%%%%

\begin{figure}
 \includegraphics[scale=0.8]{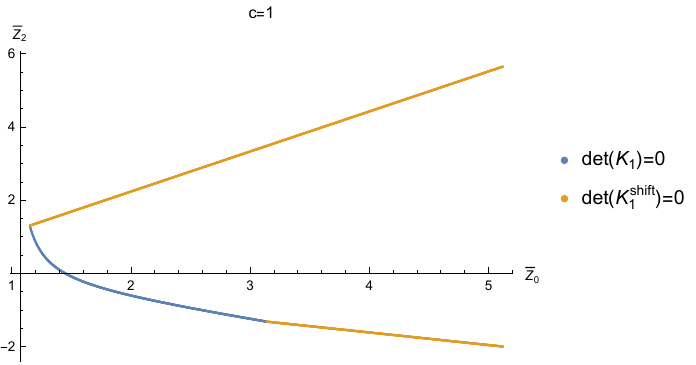}
 \centering
 \caption{The boundary of allowed region for $c=1,\Delta_{\text{gap}}=0$ of $\{\hat{Z}_0,\hat{Z}_2\}$ using $\LL=3$ constraint}
 \label{fig:z0z2}
\end{figure}

Let us being with the simplest setup, the spin-less boostrap at $\Lambda=3$ where the geometry is in $\mathbb{P}^3$. Let us consider the geometry $\{\hat{Z}_0,\hat{Z}_1,\hat{Z}_2,\hat{Z}_3\}$, where we can analytically solve the boundary. Because of modular invariance, there are the only three independent parameters $\{\M,\hat{Z}_0,\hat{Z}_2\}$, with
\begin{equation}
\hat{Z}_0=\hat{z}^{(0,0)},\quad\hat{Z}_2=2\hat{z}^{(1,1)}+\hat{z}^{(0,2)}+\hat{z}^{(2,0)}\,.
\end{equation}
%$\{\bar{Z}_0,\bar{Z}_2\}$ are also related to %$\vec{\mathcal{Z}}_3$ by a linear transformation, which can be derived from \eqref{eq: ModularHedron} taking $\tau=-\bar{\tau}=i\beta$ limit,
%\begin{align}
%&\begin{pmatrix}
%    \mathcal{Z}_0\\
%    \mathcal{Z}_1\\
%    \mathcal{Z}_2\\
%    \mathcal{Z}_3
%\end{pmatrix}
%=M_3(c,\M)\cdot
%\left[
%\begin{pmatrix}
%   \bar{Z}_0\\
%    0\\
%   \bar{Z}_2\\
%   0
%\end{pmatrix}
%-\vec{G}_{\text{vac}}\right],
%~\vec{G}_{\text{vac}}=
%\begin{pmatrix}
%\left(e^{2 \pi }-1\right)^2 e^{\frac{1}{6} \pi  (c-25)}\\
%\frac{2}{3} e^{\frac{1}{6} \pi  (c-13)} \sinh (\pi ) ((\pi  (c-13)+3) \sinh(\pi )+12 \pi \cosh(\pi))\\
%\vdots
%\end{pmatrix}\notag\\
%&M_3(c,\M)=
%\begin{pmatrix}
% 1 & 0 & 0 \\
% \frac{1}{12} \left(c-12 \M+\frac{3}{\pi }-1\right) &
%   -\frac{1}{2 \pi } & 0 \\
% \frac{\pi  (\pi  (c-12 \M-1)+6) (c-12 \M-1)+27}{144 \pi ^2} & \frac{-\pi  c+12 \pi  \M+\pi
%   -6}{12 \pi ^2} & \frac{1}{4 \pi ^2} \\
%\end{pmatrix}
%\end{align}
This space is carved out by the union of the following equation sets,
\begin{equation}\label{eqn:P3eq}
\det K_n~\text{and}~\det K_n^{{\rm twist}_\Delta}\succeq 0,\quad\text{for }n=0,1\,,
\end{equation}
where the explicit entries of the $2\times 2$ Hankel matrix $K_1=K_1[\vec{y}]$ are given by,
\begin{equation}\label{eqn:P3yvec}
\vec{y}=M(c).\left[
\begin{pmatrix}
   \hat{Z}_0\\
    0\\
   \hat{Z}_2\\
   0
\end{pmatrix}
-\vec{G}_{\text{vac}}\right]\,,
\end{equation}
where $\vec{G}_{\text{vac}}$ is the contribution from the vacuum character and $M(c)$ is the associate GL transformation that converts the character expansion into the moment form. They are explicitly given as:
\begin{align}\label{Gvac}
&\vec{G}_{\text{vac}}=
\begin{tiny}
\begin{pmatrix}
\left(e^{2 \pi }{-}1\right)^2 e^{\frac{1}{6} c_1 }\\
\frac{1}{6}\left(e^{2 \pi }{-}1\right) e^{\frac{1}{6} c_1} \left({-}c_1 {+}e^{2 \pi } (c_3{+}3){-}3\right)\\
\frac{1}{36} e^{\frac{1}{6} c_1 } \left(c^2
_1{+}12 c_1 {-}2 e^{2 \pi } \left(c^2_2{+}12 c_2{+}9\right){+}e^{4 \pi } \left(c^2_1+12 c_1+9\right)+9\right)\\
\frac{e^{\frac{1}{6} c_1}}{216} \left( c_1 ( c_1 (  c_1 {+}27){+}117){-}2 e^{2 \pi }( c_2 (  c_2 ( c_2{+}27){+}117){+}27){+}e^{4 \pi } (  c_3 ( c_3 ( c_3 {+}27){+}117){+}27){+}27\right)
\end{pmatrix}\,,
\end{tiny}
\notag\\
&M(c)=
\begin{tiny}
\begin{pmatrix}
 1 & 0 & 0 & 0 \\[2ex]
 \dfrac{c_4{+}3}{12\pi} &
   -\dfrac{1}{2 \pi } & 0 & 0 \\[2ex]
 \dfrac{  ( c_3{+}6) c_3{+}27}{144 \pi ^2} & \dfrac{- c_3{-}6}{12
   \pi ^2} & \dfrac{1}{4 \pi ^2} & 0 \\[2ex]
 \dfrac{ (  (  c_3{+}9) c_4+81) c_3{+}405}{1728 \pi ^3} & -\dfrac{ (
   c_3{+}12) c_3{+}69}{96 \pi ^3} &
   \dfrac{  c_3{+}9}{16 \pi ^3} & -\dfrac{1}{8 \pi ^3}
\end{pmatrix}
\end{tiny}\,,
\end{align}
where $c_1=\pi(c{-}25)$, $c_2=\pi(c{-}13)$ and  $c_3=\pi(c{-}1)$ respectively.  Now the boundary of the space $\{\hat{Z}_0,\hat{Z}_2\}$ is given by,
\begin{equation}\label{eqn:P3boundary}
\boxed{
\det K_1=0~\text{or}~\det K_1^{{\rm twist}_\Delta}=0
}\,.
\end{equation}
In Figure \ref{fig:z0z2} we show an example of the allowed region of $\{\hat{Z}_0,\hat{Z}_2\}$ in $c=1$ and $\Delta_{\text{gap}}=0$. First, note that the region is unbounded with the boundaries given by either $K_1$ or $K_1^{{\rm twist}_\Delta}$ being rank 1. As we will see, we can actually solve for optimal $\Delta_{\text{gap}}$ analytically in $c\to\infty$ limit.

Since we expect that beyond the gap there are no consistent solutions, i.e. the space of partition function is empty. This means that the boundaries in $\mathbb{P}^3$ becomes degenerate, $\det K_1=\det K_1^{{\rm twist}_\Delta}=0$. This condition is satisfied if there is only one state. the gap state. Then we can rewrite \eqref{eqn:P3yvec} as,
\begin{equation}
n_{\Delta_{\text{gap}}}M^{-1}(c)\cdot
\begin{pmatrix}
1\\
\Delta_{\text{gap}}\\
\Delta_{\text{gap}}^2\\
\Delta_{\text{gap}}^3
\end{pmatrix}+\vec{G}_{\text{vac}}=
\begin{pmatrix}
\hat{Z}_0\\
0\\
\hat{Z}_2\\
0
\end{pmatrix}\,.
\end{equation}
Next focus the second and fourth row of this vector equation,
\begin{equation}
n_{\Delta_{\text{gap}}}\vec{F}_{\Delta_{\text{gap}}}+\vec{F}_{\text{vac}}=
\begin{pmatrix}
0\\
0
\end{pmatrix}\,.
\end{equation}
The zero on the RHS indicates that the two vectors are collinear,
\begin{equation}\label{eqn:P3gaplargec}
\omega(\Delta_{\text{gap}})\equiv \det[\vec{F}_{\text{vac}},\vec{F}_{\Delta_{\text{gap}}}]=0\,.
\end{equation}
Now $\omega(\Delta_{\text{gap}})$ is a degree 3 polynomial in $\Delta_{\text{gap}}$, we can analyse the $c\to\infty$ behavior by first assuming $\Delta_{\text{gap}}=a_1 c+ a_0+\frac{a_1}{c}+\dots$, then solve the largest root $\Delta_*$ order by order. At leading order we have,
\begin{equation}
a_1(1+18a_1(4a_1-1))=0\Rightarrow a_1=0,\frac{1}{12},\frac{1}{6}\,.
\end{equation}
Since $\Delta_{\text{gap}}$ will be the largest zeros, $a_1=\frac{1}{6}$. Then we can continue to solve the second order equation,
\begin{equation}
-12+\pi(7+6a_0-6\coth(\pi))=0\Rightarrow a_0=\frac{12-7 \pi +6 \pi  \coth (\pi )}{6 \pi }\,.
\end{equation}
The first two order of $\Delta_{gap}$ when $c\to\infty$ are,
\begin{equation}
\Delta_{gap}=\frac{c}{6}+\frac{12-7 \pi +6 \pi  \coth (\pi )}{6 \pi }+\mathcal{O}(1/c)\,.
\end{equation}
This precisely matches with the early result in \cite{Hellerman2011-su}, where the $\mathcal{O}(c^0)$ term precisely evaluates to the numeric value 0.473695 quoted there in.

\begin{figure}[h]
  \centering
  \includegraphics[width=\textwidth]{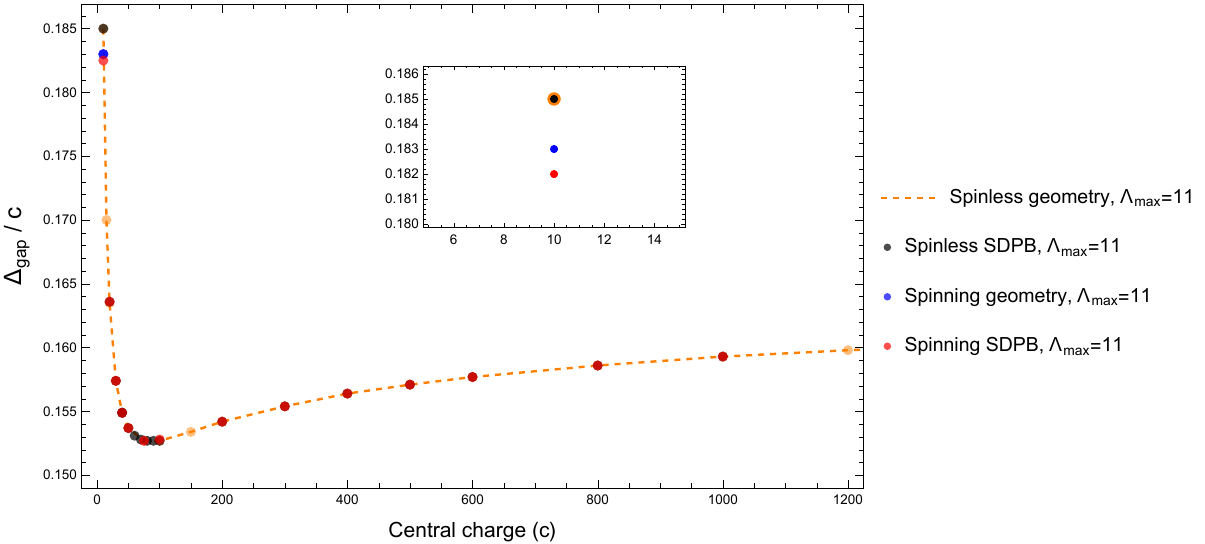}   
  \caption{Bound on \(\Delta_{\text{gap}}\) vs. central charge \(c\) for truncation order \(\Lambda_{\text{max}} = 11\). We observe that for $c>100$ there is very little difference between the four bootstraps.}
  \label{delta_gap_vs_c_1} 
\end{figure}

In Figure \ref{delta_gap_vs_c_1} we display the results for optimal gap at higher truncation order ($\Lambda_{\text{max}}=11$) with varying central charges. We compare between the geometric modular-polytope/hedron approach and SDPB. For the spinless analysis, we see that the modular polytope matches with the spinless SDPB. This is expected since for the case of a single moment, the Hankel positivities are necessary and sufficient conditions for the truncated moment problem. For the spinning analysis we see that at finite central charge, for example $c\sim 10$ in the inset of Figure \ref{delta_gap_vs_c_1}, the result of modular-hedron is slightly weaker than the spinning SDPB. Again this is reflecting that for a finite truncation order, the positivity of the Hankel matrices need not be sufficient condition for the moment geometry. However, such difference appears to be negligible at large central charges $c\gg1$. In fact, as we extend to large central charge, there is negligible difference between the spinless and spinning bootstrap. 

\begin{figure}[h]
  \centering
  \includegraphics[width=\textwidth]{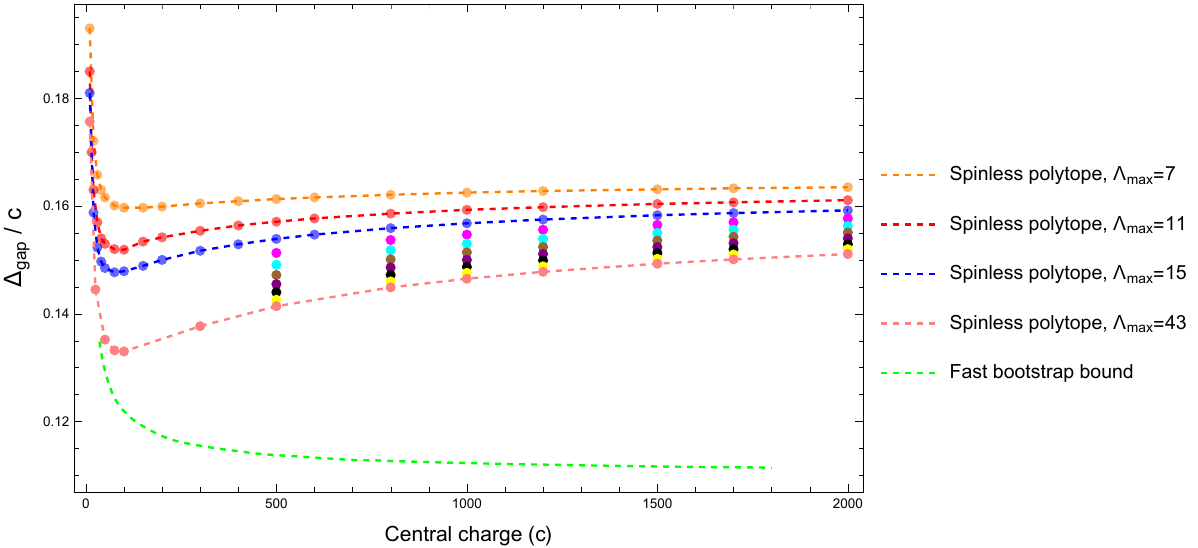}
  \caption{Bound on \(\Delta_{\text{gap}}\) vs. central charge \(c\) for different truncation orders \(\Lambda_{\text{max}}\). The color dots are the data points of truncation order $\Lambda_{max}=7,\,11,\,15,\,19,\,23,\,27,\,31,\,35,\,39,\,43.$ The fast bootstrap data are extracted from \cite{Afkhami-Jeddi2019-ti}. It is clear the bounds have not yet converged, but we are able to directly verify the $c \gtrsim  2000$ regime.}
  \label{delta_gap_vs_c_2}
\end{figure}

\begin{figure}[h]
    \centering
    \includegraphics[width=0.7\textwidth]{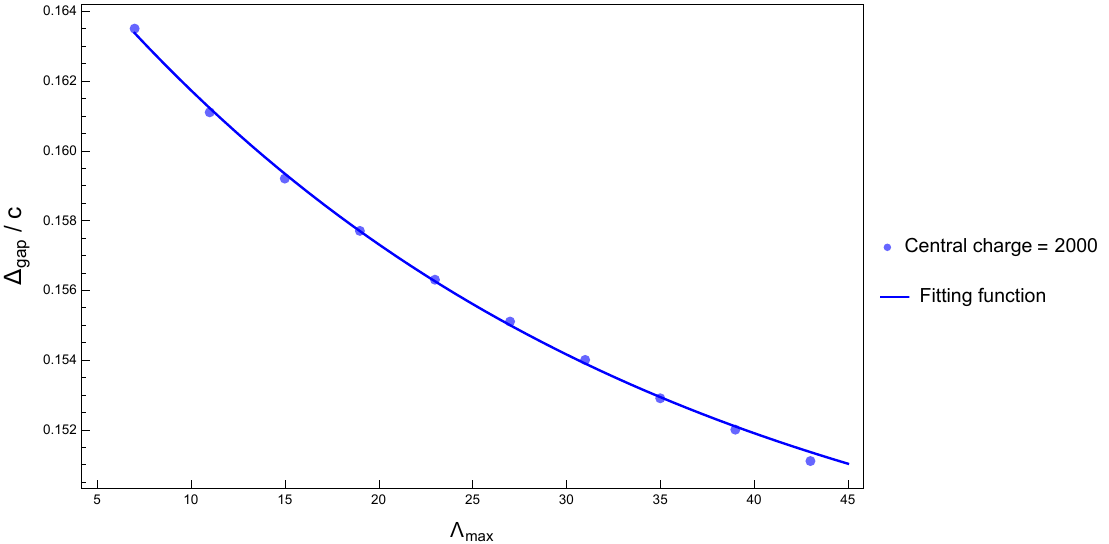}
    \caption{Bounds on $\M/c$ vs. $\Lambda_{\textrm{max}}$ at central charge equals 2000. The fitting function is obtained from $\Lambda_{max}\leq39$ data. The $\Lambda_{max}=43$ data point deviates from the fitting function. This implies that the extrapolated result has not stabilized.}
    \label{extrapolate}
\end{figure}

In Figure \ref{delta_gap_vs_c_2} we give the result for various different truncation orders up to $\Lambda_{max}=43$ and extrapolate $\Lambda_{\textrm{max}}$ to infinity with the fitting function $\M /c = A + B\, exp \left(-D \Lambda_{max}\right)$. However, we found that the extrapolated result is not stable. For instance, at central charge 2000, the extrapolated result improves when $\Lambda_{\textrm{max}}$ increases from 39 to 43. Shown in Figure \ref{extrapolate}, the fitting function derived from $\Lambda\leq39$ data deviates from the $\Lambda_{max}{=}43$ data point. This indicates that the extrapolated result has yet to converge, and it will continue to decrease as $\Lambda_{\textrm{max}}$ increases. 

\begin{table}[h]
    \centering
    \caption{The details of the fitting function.}
    \label{fitting_function_table}
\begin{tabular}{|c|c|c|}
    \hline
    Fitting Function &  $1-R^2_{adj}$ & $A$ \\
    \hline
    $A+B/c+D/c^2$ & $10^{-7}$ & $1/6.60$ \\
    \hline
    $A+B/c+D \,log(c)/c$ & $4\times10^{-8}$ & $1/6.44$\\
    \hline
\end{tabular}
\end{table}
We continue to fit the extrapolated result and obtain an estimation of $\M/c$ at the limit $c\rightarrow\infty$. We fit the data within the range $500 \leq c \leq 2000$. The details of the fitting functions are described in Table \ref{fitting_function_table} which is also used in \cite{Afkhami-Jeddi2019-ti}. The estimation differs since different functions were used. However, both fitting functions estimated that the ratio $\M/c$ is approximately $1/6.60 \sim 1/6.44$ at the infinite $c$ limit. We expect that the result will improve as we increase the $\Lambda_{\textrm{max}}$ truncation. \\

\begin{figure}[h]
  \includegraphics[width=\textwidth]{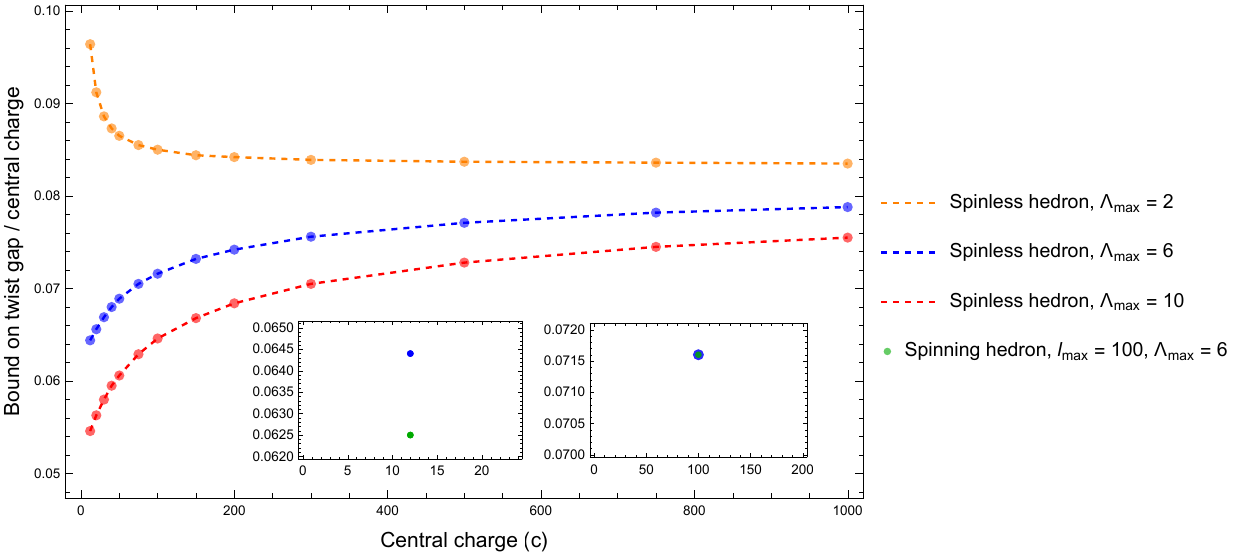}
  \centering
  \caption{Bound on the twist gap vs. central charge \(c\) at different derivative orders.}
  \label{twistGapPlt}
\end{figure}

\noindent \textbf{Bounding the twist-gap} The twist-gap corresponds to ${\rm min}[h,\bar{h}]\geq h_{\text{gap}}$, which means \(h \geq h_{\text{gap}}\) and \(\bar{h} \geq h_{\text{gap}}\). Translated to our geometry, one modifies the Hankel in eq.(\ref{constraint_2}) to the twisted Hankel $\mathbf{K}[\hat{\mathbf{Z}}_{h_{gap}}]$ by identifying 
\begin{equation}
  \mathbf{K}[\hat{\mathbf{Z}}_{h_{gap}}]^{\text{twist}_{h}} = \sum_{h, \bar{h}} \tilde{n}_{h, \bar{h}} (h {-}h_{gap})
  \begin{pmatrix}
    1 \\ h \\ \bar{h} \\ \vdots
  \end{pmatrix}
  \begin{pmatrix}
    1 \\ h \\ \bar{h} \\ \vdots
  \end{pmatrix}^T \succeq 0\,.
\end{equation}
Similarly, to impose \(\bar{h} \geq h_{\text{gap}}\), 
\begin{equation}
  \mathbf{K}[\hat{\mathbf{Z}}_{h_{gap}}]^{\text{twist}_{h}} = \sum_{h, \bar{h}} \tilde{n}_{h, \bar{h}} (\bar{h} {-} h_{gap})
  \begin{pmatrix}
    1 \\ h \\ \bar{h} \\ \vdots
  \end{pmatrix}
  \begin{pmatrix}
    1 \\ h \\ \bar{h} \\ \vdots
  \end{pmatrix}^T \succeq 0\,.
\end{equation}
We demonstrate the results in Figure \ref{twistGapPlt}, again for various central charges. Here, we compare the result from the modular-hedron geometry with or without integer spin constraints. As evident in the plot, the difference is negligible when  $c>10$.  At $c=100$ the difference is less than $0.1\%$, so for sufficiently large central charge, spin integrality will not significantly improve the bounds.

%%%%%%%%%%%%%%%%%%%%%%%%%%%%%%%%%%%%%%%%%%%
\subsection{Two sided bounds for $Z_0$ and critical gap}
%%%%%%%%%%%%%%%%%%%%%%%%%%%%%%%%%%%%%%%%%%%

So far, we've demonstrated how to extract the optimal gap or twist gap from our geometric approach. It is also interesting to focus on the region of intersection itself, representing the ``space" of partition functions. For this purpose it will be sufficient to consider the modular-polytope $\vec{\mathcal{Z}}^{\rm spinless}$ eq.(\ref{eq: ModularPoly}). Once again we simply have that the Hankel and twisted-Hankel must be positive:
\eq
        \mathbf{K}[\vec{\mathcal{Z}}^{\rm spinless}] \succeq 0, \quad \mathbf{K}[\vec{\mathcal{Z}}^{\rm spinless}]^{\text{twist}_{\Delta}} \succeq 0\,.
\eqe
First we claim that $\mathcal{Z}^{\rm spinless}_q$ is bounded from both sides. Recall that modular invariance implies that $\hat{z}^{(m,n)}=0$ and $m{+}n=odd$. Due to the GL transformation, $\mathcal{Z}^{\rm spinless}_q$ for $q=odd$ will be linear combination of $\mathcal{Z}^{\rm spinless}_i$ with $i<q$.  Now, starting with $\mathcal{Z}^{\rm spinless}_0$, every time we introduce two more derivatives to the partition function, the Hankel conditions are also enlarged to include two new inequalities linear/quadratic in the new variable $\mathcal{Z}^{\rm spinless}_n$, respectively. The highest unshifted Hankel determinant reads
\[
\det
\begin{pmatrix}
 K_{n-1}&a\\a^T&\mathcal{Z}^{\rm spinless}_n
\end{pmatrix}\geq0\,,
\]
which means the term linear in the new variable $\mathcal{Z}^{\rm spinless}_n$ is non-negative. 
On the other hand, the highest twisted Hankel determinant reads
\[
\det
\begin{pmatrix}
 K^{\text{twist}_{\Delta}}_{n-2}&*&*\\
 *&*&\mathcal{Z}^{\rm spinless}_n \\ * &\mathcal{Z}^{\rm spinless}_n & \mathcal{Z}^{\rm spinless}_{n{+}1}
\end{pmatrix}\geq0\,.
\]
This gives a quadratic inequality whose leading coefficient for $(\mathcal{Z}^{\rm spinless}_n)^2$ is $-K^{\text{twist}_{\Delta}}_{n{-}2}$, which is non-positive. Thus the former condition gives a lower bound for $\mathcal{Z}^{\rm spinless}_n$, while the later an upper bound. 

\begin{figure}
    \centering
  \begin{subfigure}[b]{0.48\textwidth}
         \centering
   \includegraphics[width=1\textwidth]{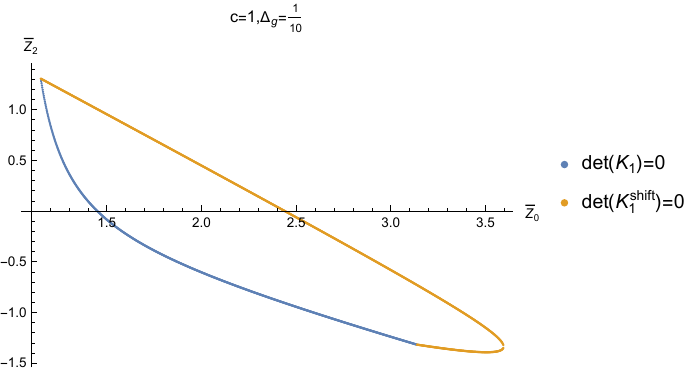}
   \end{subfigure}
   \begin{subfigure}[b]{0.48\textwidth}
         \centering
   \includegraphics[width=1\textwidth]{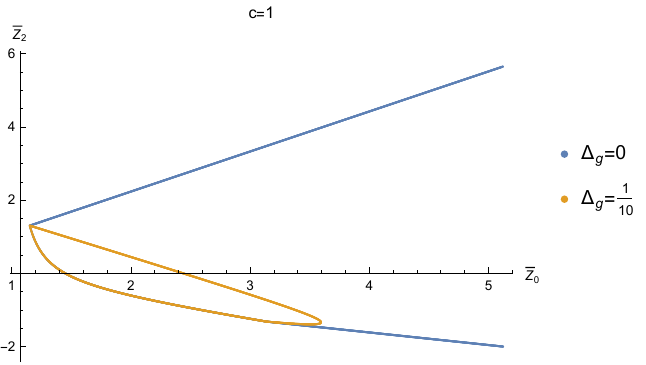}
   \end{subfigure}
    \caption{Left: The boundary of allowed region for $c=1,\Delta_{\text{gap}}=\frac{1}{10}$ of $\{\hat{Z}_0,\hat{Z}_2\}$ space, at $\LL=3$.  Right: Comparison of $\Delta_{\text{gap}}=0$ and $\Delta_{\text{gap}}=\frac{1}{10}$ in $\{\hat{Z}_0,\hat{Z}_2\}$ space when $c=1$. }
    \label{fig:z0z2gap}
\end{figure}

Therefore the boundedness of the space boils down to $\mathcal{Z}^{\rm spinless}_0$, which is always bounded from below since it is definite non-negative. As we will see there is a critical gap $\Delta_{\text{gap}}^*$ above which $\mathcal{Z}^{\rm spinless}_0$ develops an upper bound. Once again let us consider the $\mathbb{P}^3$ modular polytope geometry in more detail. The constraints $\det K_0^{\text{twist}_{\Delta}}\geq0$ and $\det K_0\geq0$ in \eqref{eqn:P3eq} take the form 
\begin{align}\label{eqn:K1gap}
&\det K_0\geq0\Rightarrow  \hat{Z}_0\geq\left(e^{2 \pi }-1\right)^2 e^{\frac{1}{6} c_1}\,,\notag\\
&\det K_0^{\text{twist}_{\Delta}}\geq0\Rightarrow\frac{ \hat{Z}_0 (\pi  c_4{+}3)}{12 \pi }+e^{\frac{1}{6} c_1} (\Delta_{\text{gap}}-2)-2e^{\frac{1}{6} c_2} (\Delta_{\text{gap}}-1)+e^{\frac{1}{6} c_3}\Delta_{\text{gap}}\geq0\,.
\end{align}
From \eqref{eqn:K1gap}, we see that when the coefficient of $\hat{Z}_0$ turns negative the constraint becomes an upper bound on $\hat{Z}_0$. This occurs when the gap is above a critical value, 
\begin{equation}\label{eq:z0upperp3}
\boxed{
\Delta_{\text{gap}}\geq \frac{c-1}{12}+\frac{1}{4\pi}
}\,.
\end{equation}
In this case, $\hat{Z}_0$ will be bounded from two sides, 
\begin{equation}\label{eqn:z0twosidedbound2}
e^{\frac{c_1}{6}}(e^{2\pi}-1)^2\leq\hat{Z}_0\leq\frac{12 \left(e^{2 \pi }-1\right) \pi  e^{\frac{1}{6} c_1}
   \left(\left(e^{2 \pi }{-}1\right)\Delta_{\text{gap}}{+}2\right)}{{-}\pi c_4 {-}3}\,.
\end{equation}
Now let's move to other constraints $\det K_1\geq0$ and $\det K_1^{\text{twist}_{\Delta}}\geq0$ in \eqref{eqn:P3eq},
\begin{align}\label{eqn:P3K3eq}
&36 \left(\hat{Z}_0-\left(e^{2 \pi }{-}1\right)^2 e^{\frac{1}{6} c_1}\right)\hat{Z}_2-e^{\frac{1}{6}c_1} (c_1 (c_1+6)+27) \hat{Z}_0\notag\\
&+2e^{\frac{1}{6} c_2} (c_2 (c_2+6)+27)
   \hat{Z}_0-e^{\frac{1}{6} c_3} (c_3 (c_3+6)+27)\notag\\
&+ \hat{Z}_0-288 \pi ^2 e^{\frac{1}{3} \pi  (c-19)}+576 \pi ^2
   e^{\frac{1}{3} c_2}-288 \pi ^2 e^{\frac{1}{3} \pi  (c-7)}+18
    \hat{Z}_0^2\geq0\,,\notag\\
& - \hat{Z}_2^2+f_1(\hat{Z}_0)\hat{Z}_2+f_0(\hat{Z}_0)\geq0\,,
\end{align}
where $\det K_1$ is linear and $\det K_1^{\text{twist}_{\Delta}}$ is quadratic in $\hat{Z}_2$. The lower bound for $\hat{Z}_0$ in \eqref{eqn:z0twosidedbound2} ensures that the coefficient for $\hat{Z}_2$ is positive in the first inequality above, implying a lower bound.  From the second inequality,  the discriminant for the quadratic polynomial in $\hat{Z}_2$ is always positive as long as $\hat{Z}_0$ satisfies \eqref{eqn:z0twosidedbound2}, leading to an upper bound for $\hat{Z}_2$. We see that the space of $\{\hat{Z}_0,\hat{Z}_2\}$ is finite once the gap of the spectrum satisfies \eqref{eq:z0upperp3}. Figure \ref{fig:z0z2gap} shows an example at $c=1,\Delta_{\text{gap}}=1/10$. As we increase $\M$, the region shrinks, until it vanishes completely when $\M$ reaches the maximal allowed value.
%And the boundary is still given by,
%\begin{equation}
%\text{Upper Bound:}~\det K_3^{\text{shift}}=0,\quad\text{Lower Bound:}~\det K_3=0
%\end{equation}

\paragraph{Critical gap and HKS bound}The presence of a \emph{critical gap}, above which the partition function is bounded, can be derived from the Hartman, Keller, Stoica \cite{Hartman:2014oaa,Mukhametzhanov:2019pzy} (we will call it HKS bound from now on) bound. To see this, we start with (6.6) in \cite{Mukhametzhanov:2019pzy}\footnote{The $\beta$ here is different than the $\beta$ in (6.6) in \cite{Mukhametzhanov:2019pzy} by $2\pi$, and we are using the reduced partition which subtracted the Dedekind eta functions. So the exact form of \eqref{eqn:HKS} will be different from (6.6) in \cite{Mukhametzhanov:2019pzy}.}
\begin{equation}\label{eqn:HKS}
Z[\beta]<Z_L(1/\beta)+\frac{Z_L[1/\beta]-Z_L[\beta]}{\beta e^{\pi/6(\beta-1/\beta)(c-1-12\Delta_H)}-1},\quad(\Delta_H>\frac{c-1}{12},\beta\geq1)\,,
\end{equation}
where $Z[\beta]$ is defined in \eqref{eqn:spinlessZ} and $Z_L(\beta)$ comes from summing over all the contributions from states having scaling dimensions below $\Delta_H$,
\begin{equation}
Z_L[\beta]=\beta^{1/2}e^{2\pi\beta\frac{c-1}{12}}(1-e^{-2\pi\beta})^2+\sum_{\Delta\leq\Delta_H} n_\Delta e^{2\pi\beta\Big(\Delta-\frac{c-1}{12}\Big)}\,.
\end{equation}
Here the separation between low and high energy states can be arbitrary as long as the threshold $\Delta_H$ is larger than $\frac{c-1}{12}$. Now we take the self-dual point limit $\beta\to1$. \eqref{eqn:HKS} becomes,
\begin{equation}\label{eqn:HKSselfdual}
\hat{Z}_0\leq Z_L[1]-\frac{Z_{L}^{'}[1]}{3+\pi(c-1-12\Delta_H)},\quad \Delta_H>\frac{c-1}{12}\,.
\end{equation}
Without knowing the low energy spectrum, we cannot guarantee a upper bound from \eqref{eqn:HKSselfdual} since one can always assume a state with small scaling dimension and large degeneracy. In such case the upper bound of $\hat{Z}_0$ can become arbitrarily large. Now let's consider the case where the first operator $\M$ in CFT has scaling dimension larger than $\frac{c-1}{12}$ and we set $\Delta_H=\M$ in \eqref{eqn:HKSselfdual}. Then the RHS will only receive contribution from the vacuum character,
\begin{equation}\label{eqn:HKSselfdualfinal}
\hat{Z}_0\leq\frac{12 \left(e^{2 \pi }-1\right)
   \pi  e^{\frac{1}{6} \pi  (c-25)}
   \left(\left(e^{2 \pi }-1\right)
   \Delta _{\text{gap}}+2\right)}{-\pi  c+12
   \pi  \Delta _{\text{gap}}+\pi -3},\quad\Delta_{\text{gap}}=\Delta_H>\frac{c-1}{12}\,.
\end{equation}
In order for this to be a valid upper bound, the denominator in \eqref{eqn:HKSselfdualfinal} must be positive, which implies a critical gap $\Delta^*_{\text{gap}}$
\begin{equation}
\Delta^*_{\text{gap}}=\frac{c-1}{12}+\frac{1}{4\pi}\,.
\end{equation}
We see that from the HKS bound \eqref{eqn:HKS} we extract an upper bound for $\hat{Z}_0$, 
\begin{equation}\label{eqn:HKSselfdualfinal2}
\boxed{
\hat{Z}_0\leq\frac{12 \left(e^{2 \pi }-1\right)
   \pi  e^{\frac{1}{6} \pi  (c-25)}
   \left(\left(e^{2 \pi }-1\right)
   \Delta _{\text{gap}}+2\right)}{-\pi  c+12
   \pi  \Delta _{\text{gap}}+\pi -3},\quad\Delta_{\text{gap}}\geq \Delta^*_{\text{gap}}
   }\,,
\end{equation}
with $\Delta^*_{\text{gap}}=\frac{c{-}1}{12}{+}\frac{1}{4\pi}$. This exactly matches with our result in \eqref{eqn:z0twosidedbound2} and \eqref{eq:z0upperp3} respectively.

\begin{figure}[h]
  \centering
  \includegraphics[width=0.8\textwidth]{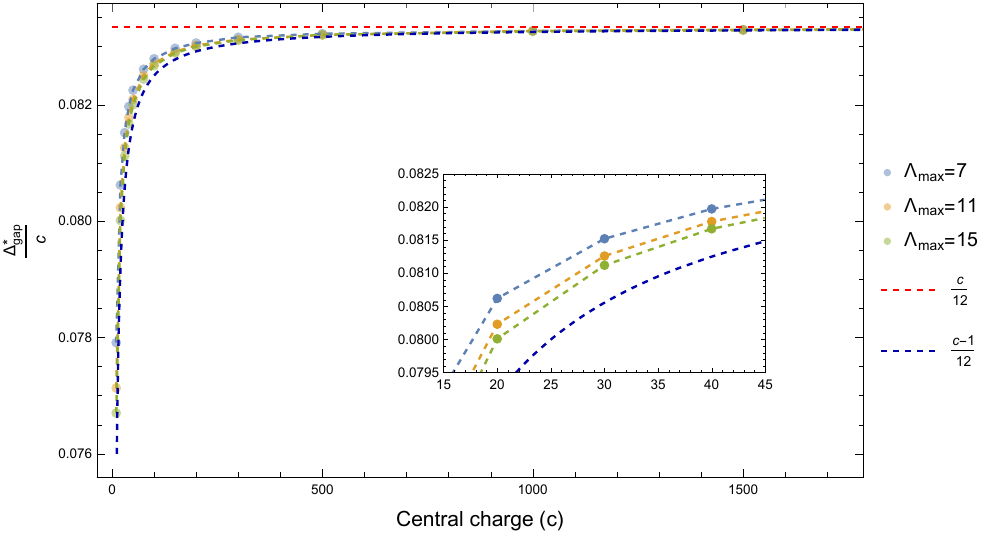}
  \caption{Critical gap \(\Delta_{\text{gap}}^{*}\) vs. central charge \(c\). Numerical results at large $c$ suggest an asymptotic convergence of the critical gap toward $c/12$, and we observe that the critical gaps at various derivative orders alway lies above $\frac{c-1}{12}$.}
  \label{crit_gap_vs_c}
\end{figure}

While the bound derived from HKS matches with our analytic result at $\Lambda_{max}=3$, we expect the bounds to improve significantly as we increase $\Lambda$. To numerically determine the critical gap, we decrease the value of \(\Delta_{\text{gap}}\) until no upper bound on $\hat{Z}_0$ can be found. A more detailed discussion on the method of certifying unboundedness is given in Appendix \ref{appendix_SDP}. In Figure \ref{crit_gap_at}, the critical gap at various different derivative orders for $c = 12$ and $c = 100$ are considered. Using $\Delta_{\text{gap}}^* / c = a + b \Lambda_{\text{max}}^{-p}$ as an extrapolation function, we extrapolate the infinite $\Lambda$ limit giving  $\Delta_{\text{gap}}^* / c \approx 0.07606$ and $\Delta_{\text{gap}}^* / c \approx 0.08247$ for $c = 12$ and $c = 100$, respectively, both being nonzero. Note that for $c = 100$ the $\Delta_{\text{gap}}^* / c$ is approximately close to $\frac{1}{12}$. We plot the value of  critical gap against the central charge, shown in Figure \ref{crit_gap_vs_c}, which shows that $\Delta_{\text{gap}}^{*}$ lies between $\frac{c{-}1}{12}$ and $\frac{c}{12}$, and  converges towards $\frac{c}{12}$ at large $c$.

The fact that partition function is bounded from above for theories with gap above $\frac{c{-}1}{12}$ is closely connected to the degeneracy being bounded. Indeed as shown in \cite{Collier:2016cls}, for gap above $\frac{c{-}1}{12}$ and below the optimal gap at a given derivative order, it is possible to find a functional for the modular constraint that is positive above the gap and only negative on the identity block. The ratio between the functional evaluated at the identity and a particular conformal dimension then gives the upper bound of the degeneracy number for that state. The lower limit $\frac{c{-}1}{12}$ reflects the existence of Liouville theory which has a non-normalizable vacuum, hence effectively unbounded degeneracy number.

\begin{figure}
    \centering
    \begin{subfigure}[h]{0.49\textwidth}
        \includegraphics[width = \textwidth]{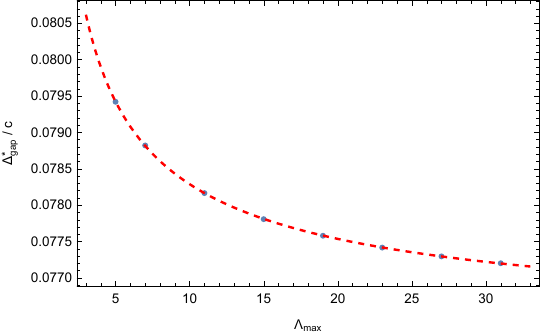}
        \caption{$c = 12$}
    \end{subfigure}
    \begin{subfigure}[h]{0.49\textwidth}
          \includegraphics[width = \textwidth]{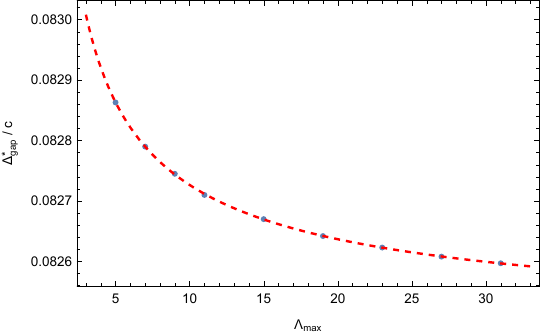}
          \caption{$c = 100$}
    \end{subfigure}
    \caption{Estimation of the critical gap, where the fitting function $\Delta_{\text{gap}}^* / c = a + b \Lambda_{\text{max}}^{-p}$ is used. $a = 0.076063$ is obtained for $c = 12$, and $a = 0.082466$ for $c = 100$. In both cases we have $p \approx 0.6$.}
    \label{crit_gap_at}
\end{figure}

\begin{figure}[h]
  \centering
  \includegraphics[width=0.8\textwidth]{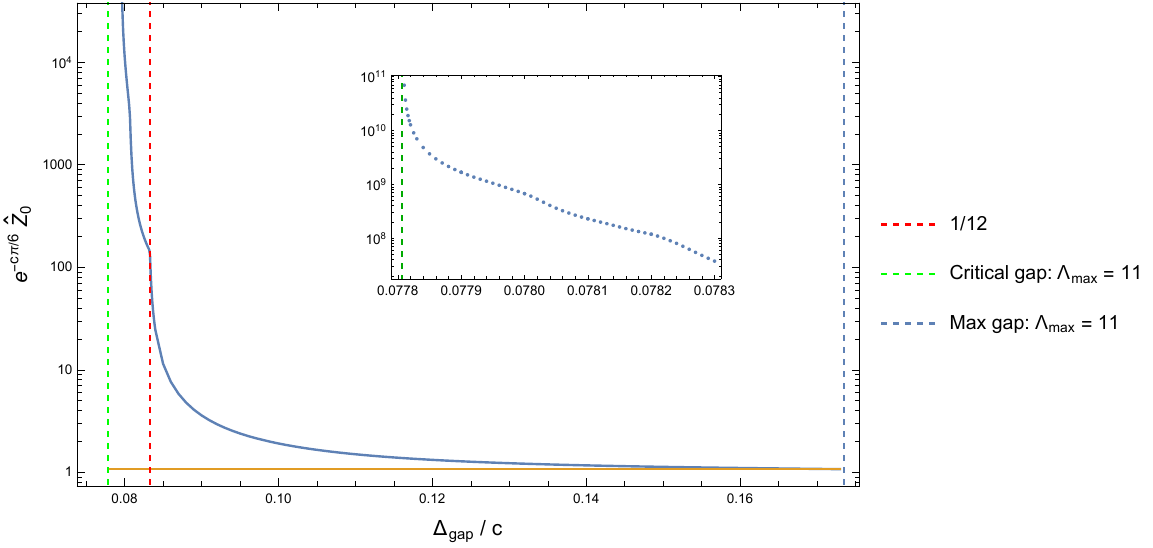}
  \caption{Upper and lower bounds on $Z^{(0,0)}$ plotted against the gap for $c = 12$. The space vanishes when reaching the maximal gap. The upper bound grows super-exponentially, becoming infinite at the critical gap. Notice the kink near \(\Delta_{\text{gap}} = \frac{c}{12}\), which divides the allowed region of $\Delta_{\text{gap}}$ into two parts. }
  \label{determine_crit_gap}
\end{figure}

\begin{figure}[H]
  \centering
  \includegraphics[width=1\textwidth]{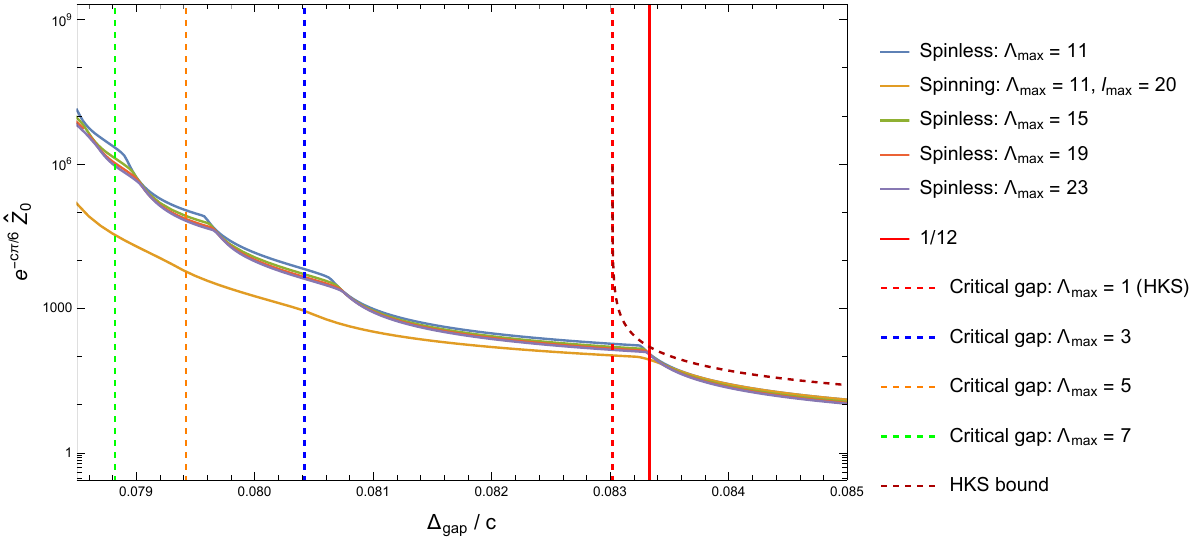}
  \caption{The upper bound on $\hat{Z}_{0}$ plotted against the gap for $c = 12$. }
  \label{kinks_c_12}
\end{figure}

\begin{figure}[H]
  \centering
  \includegraphics[width=1\textwidth]{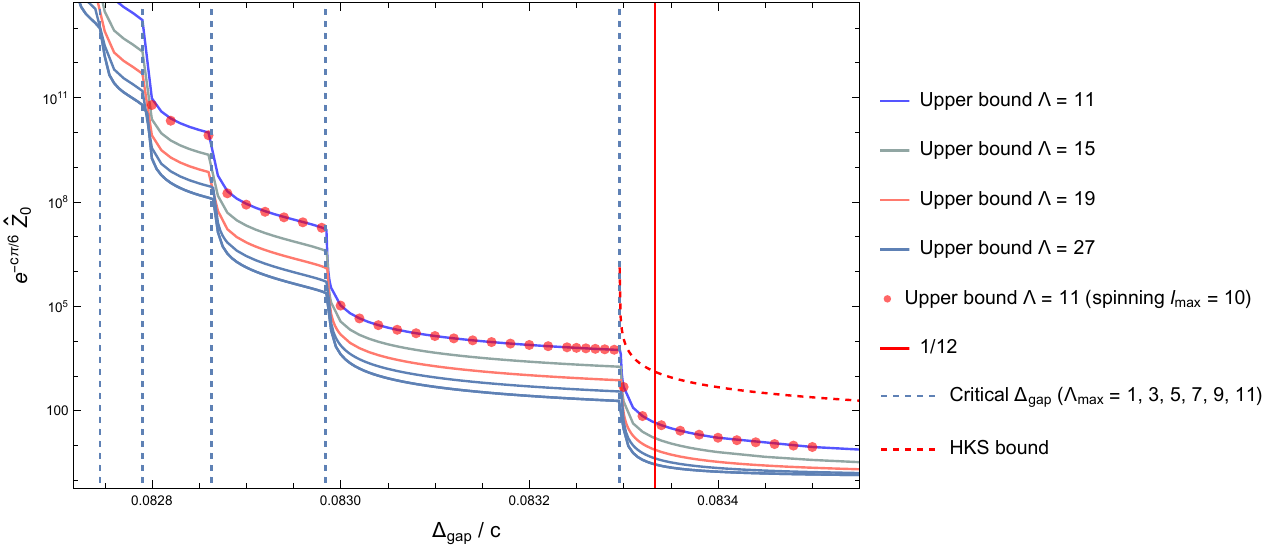}
  \caption{The upper bound on $\hat{Z}_{0}$ plotted against the gap for $c = 100$. There are multiple kinks, situated close to the lower-derivative order critical gaps.}
  \label{kinks_c_100}
\end{figure}

\paragraph{Kinks between $\frac{c}{12}$ and the critical gap}Interestingly, for sufficiently large $c$ the allowed region of $\Delta_{\text{gap}}$ can be split into two parts depending on the behaviour of $\hat{Z}_0$'s upper bound: (I) $\Delta_{\text{gap}}^* \sim c/12$ and (II) $c/12 \sim \max\{\Delta_{\text{gap}}\}$. In region (II) the upper bound varies smoothly with $\Delta_{\text{gap}}$, while in region (I) we observe the presence of \emph{kinks}, where the slope of the curve develops discontinuities as shown in Figure \ref{determine_crit_gap}. Zoomed-in plots for $c = 12$ and $c = 100$, shown in Figure \ref{kinks_c_12} and \ref{kinks_c_100}, show that kinks can be found around the lower derivative critical gaps. For $c = 12$, the upper bound of $\hat{Z}_0$ is more stable against $\Lambda_{\text{max}}$, but the kinks shift slightly. On the other hand, for $c = 100$, the upper bound decreases significantly as we increase $\Lambda_{\text{max}}$, but the values of $\Delta_{\text{gap}}$ where the kinks are located are relatively stable. 

For $c = 12$, the upper bound is more sensitive to the integer-spin constraints. As seen in Figure \ref{kinks_c_12}, the spinning bootstrap yields stronger result compared with the spinless bootstrap even at lower derivative order. Imposing the spinning constraints, the sharp kinks turn into smooth bumps. However, for large central charge, imposing integer-spin constraints does not significantly change the upper bound of $\hat{Z}_0$, and the kinks survive. The phenomenon that higher central charge is insensitive to the integer spin constraint is consistent with our earlier observation on bounding the gap at large central charge. 

In summary, we find that the the space of partition functions is bounded once the gap of the spectrum is above the critical gap. The value of critical gap lies between $\frac{c{-}1}{12}$ and $\frac{c}{12}$, and we observe the presence of stable kinks emerging which are stable under increase of derivative truncation. If we conjecture that the stable kinks point to physical theories, the fact that they only appear below $c/12$ for sufficiently large central charge is consistent with the expectation that the physical theories can only admit $\Delta_{\text{gap}} \leq c/12$ in the large $c$ limit expected from holography.

% \begin{figure}[h]\label{stabalizing_bound}
%  \centering
%   \includegraphics[width=0.8\textwidth]{c10FirstKink.png}
%   \caption{A zoomed-in version of the plot above near the kink. As the truncation order $\Lambda$ increases, the kink shifts but remains, with its position converging toward a point very stable against $\Lambda$.}
% \end{figure}

\section{Integer degeneracy number}\label{section5}

%e^{(c-1-24\Delta_i)\frac{\pi}{24}}

    So far, we have only considered the convex geometry implied by imposing the positivity of the degeneracy number. In this section we explore implications of requiring the degeneracy number to be an integer. We will solve this problem geometrically by using the method introduced in \cite{Chiang:2022jep,Chiang:2022ltp}, which constructs the allowed space iteratively via Minkowski sums of all possible states contributing to the partition function. 
    
    Returning to eq.(\ref{reduceS})
\eq
{\hat{z}}^{(p,q)}=\hat{\chi}_0^{(p)}\hat{\bar{\chi}}_0^{(q)}+\sum_i n_{h_i,\bar{h}_i} \hat{\chi}_{{h}_i}^{(p)}\hat{\bar{\chi}}_{\bar{h}_i}^{(q)}\,,
\eqe
we now require $n_{h_i,\bar{h}_i}\in \mathbb{N}$. We will focus on the spinless problem in $\Delta$ space, which amounts to studying:
\eq\label{spinlessint}
{\hat{Z}}_{k}=G_{vac}^{(k)}+\sum_i n_{i} \hat{\chi}_{i}^{(k)},\quad  n_{\Delta_i}\in \mathbb{N}\,,
\eqe
where  $\hat{Z}_{k}=\sum_{p+q=k} \left(\begin{array}{c}k \\ q\end{array}\right) \hat{Z}^{(p,q)}$, and similar for the $\hat{\chi}^{(k)}$. Explicitly $G_{vac}$ is given in eq.(\ref{Gvac}), and 
\eq
\hat{{\bm \chi}}_i=e^{(c-1-12\Delta_i)\frac{\pi}{6}}\begin{pmatrix}1\\
\frac{1}{6}(3+X)\\
\frac{1}{6^2}(9+12X+X^2)\\
\frac{1}{6^3}(27+117X+27X^2+X^3)\\
\vdots
\end{pmatrix}\,,
\eqe
where $X=( c-1 - 12 \Delta_i) \pi$. Unlike the case where $n_i$ is simply assumed to be positive, the exponential factor $e^{(c-1-12\Delta_i)\frac{\pi}{6}}$ can no longer be scaled away if we wish to impose $n_i\in \mathbb{N}$.  

Our first goal is to find the boundaries of the space $(\hat{Z}_0,\hat{Z}_1,\hat{Z}_2,\ldots)$. We focus on the space of consecutive couplings, as only in this space the boundaries have no non-trivial self intersections. Once we have the boundaries, all that remains is computing their intersection with the modular planes, $\hat{Z}_{odd}=0$. The intersection with one plane ($\Lambda_{max}=2$) can remarkably be solved in closed form, and we are able to also numerically solve the intersection with two null planes ($\Lambda_{max}=3$). We will find that already these results are in some regions stronger even than the spinning bootstrap at large $\Lambda_{\textrm{max}}$. Intersections with even more null planes would impose even stronger constraints, and can still be carried out using our methods, but computation time may be high.

\subsection{A Minkowski sum of curves}
First let us briefly explain how the allowed space of $\hat{Z}_k$ can be viewed as a Minkowski sum of curves and introduce some notation.  Slightly rewriting eq.(\ref{spinlessint}), define 
\eq\label{Adef}
 {\bf A}(n_i)(\Delta_i)\equiv {\bf G}_{vac}+n_0 \hat{\bm {\chi}}(\M)+\sum_{i=1}^\infty n_i \hat{\bm {\chi}}(\Delta_i)\,,
\eqe
assuming wlog $\Delta_j\ge \Delta_i\ge \M$ for $j\ge i$. Any given configuration of $n_i$ and $\Delta_i$ corresponds to a particular point in partition function space, and so the allowed space for $\hat{Z}_k$, for some given gap $\M$, is simply the image of $\m_i\ge \M$ and $n_i\in \mathbb{N}\, $\footnote{If any particular state $i$ is known to be present (typically the gap state), we should also require $n_i\ge 1$. We will return to this point later.} under eq.(\ref{Adef}). Our task is therefore to compute this image, and in particular to find its boundaries. 

It is in fact straightforward to show that the problem reduces to finding boundaries only for configurations of the type 
\eq
 {\bf A}_N\equiv  {\bf A}(0,\underbrace{1,1,\ldots,1}_{N})={\bf G}_{vac}+\sum_{i=1}^N  \hat{\bm {\chi}}(\Delta_i)\,.
\eqe
The crucial observation is that any ${\bf A}(n_i)(\Delta_i)$ can be obtained from some ${\bf A}(0,1,1,1,\ldots)(\Delta'_i)$ by simply fixing the $\Delta'_i$ appropriately. For instance
\eq\label{image}
{\bf A}(n_0,n_1,0,\ldots)(\Delta_1,\ldots)={\bf A}(0,\underbrace{1,\ldots,1}_{n_0},\underbrace{1,\ldots,1}_{n_1},1,\ldots)(\underbrace{\M,\ldots,\M}_{n_0}\underbrace{\Delta_1,\ldots,\Delta_1}_{n_1},\infty,\ldots)\,.
\eqe
States with large $\Delta$ (in general, with $\Delta{\gg} c$) have vanishing contribution to the partition function. For $N\rightarrow \infty$, the converse is immediately also true.
Therefore we conclude that the space ${\bf A}(n_i)(\Delta_i)$ is equivalent to the space ${\bf A}_{N\rightarrow \infty}$. We will obtain the space for ${\bf A}_\infty$ by constructing ${\bf A}_N$ iteratively, and taking the limit $N\rightarrow \infty$.

Next we discuss how the image of $\Delta_i$ may be computed. 
  
\subsubsection{Two dimensions}
\begin{figure}[H]
  \centering
  \begin{subfigure}[b]{0.48\textwidth}
         \centering
  \includegraphics[height=1.8in]{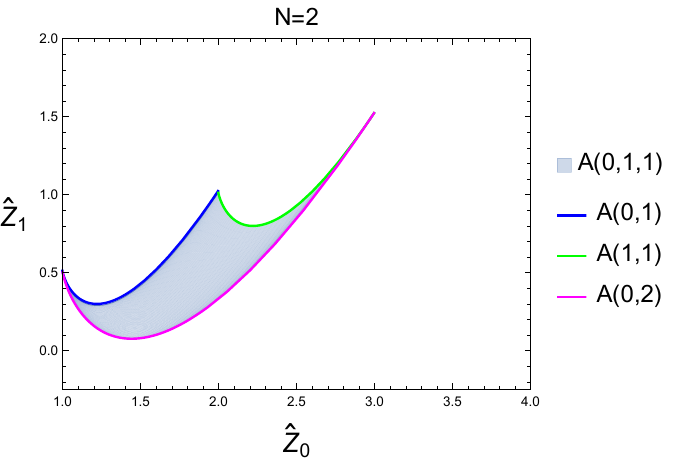} 
  \caption{${\bf A}_2$}
  \label{F0a}
  \end{subfigure}
  \begin{subfigure}[b]{0.48\textwidth}
         \centering
    \includegraphics[height=1.8in]{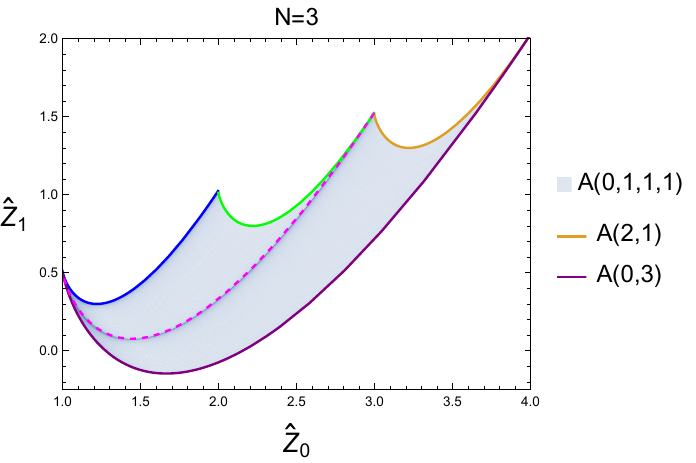} 
     \caption{${\bf A}_3$}
      \label{F0b}
    \end{subfigure}
  \caption{Allowed space of two states ${\bf A}_2$ and respectively three states ${\bf A}_3$. Note ${\bf A}_2$ is a subspace of ${\bf A}_3$. We have fixed $c=1$ for this figure.}
  \label{F0}
\end{figure}

Let us begin with the 2D space $(\hat{Z}_0,\hat{Z}_1)$, and $N=1$, ie. the space of just one state satisfying $\Delta\ge 0$, plus the vacuum. We have
\eq
{\bf A}_1={\bf A}(0,1)={\bf G}_{vac}+\hat{\bm {\chi}}(\Delta_1)=\begin{pmatrix}
G_{vac}^{(0)}+\hat{\chi}_{\Delta_1}^{(0)}\\
G_{vac}^{(1)}+\hat{\chi}_{\Delta_1}^{(1)}
\end{pmatrix}\,,
\eqe
This space is a curve, paramaterized by $ \Delta_1\ge 0$, which we plot in Figure \ref{F0} as the blue curve.  For $N=2$ we now have
\eq
{\bf A}_2={\bf A}(0,1,1)={\bf G}_{vac}+\hat{\bm {\chi}}(\Delta_1)+\hat{\bm {\chi}}(\Delta_2)\,.
\eqe
Since we need the image of both $\Delta_1$ and $\Delta_2$, this is simply the Minkowski sum of the previous curve ${\bf A}_1$ with another curve $\hat{\bm {\chi}}$. The result is a 2D surface, parameterized by the two $\Delta_i$. It is straightforward to plot the space, and one finds a 2D region with three boundaries, given by ${\bf A}(0;1)$, ${\bf A}(1;1)$ and ${\bf A}(0;2)$. This is shown in Figure \ref{F0a}. 

In general, boundaries of Minkowski sums can be of two types. The first type  are associated to bounds of the parameters themselves. Indeed  ${\bf A}(0,1)$ can be obtained from ${\bf A}(0,1,1)$  by taking $\m_2=\infty$, and ${\bf A}(1,1)$ can be obtained by taking $\m_1=0$. 
The second type of boundaries can be obtained by extremizing in one direction, keeping the other directions fixed. This can done via Lagrange multipliers, which in 2D entails solving\footnote{One can also compute the bordered Hessian to establish whether a boundary is upper or lower, but we do no find it necessary in our low dimension examples.}  
\eq
\begin{pmatrix}
\partial_{\m_1}A^{(0)}(0,1,1)&\partial_{\m_2}A^{(0)}(0,1,1)\,,\\
\partial_{\m_1}A^{(1)}(0,1,1)&\partial_{\m_2}A^{(1)}(0,1,1)\,,
\end{pmatrix}\propto (\m_1-\m_2)=0\,.
\eqe
We find the extremal solution $\m_1=\m_2$, corresponding to the third boundary ${\bf A}(0,2)$. In fact, even in higher dimensions and arbitrary number of curves, solutions obtained via extremizing will always be of the type $\m_i=\m_j$, ensuring remarkably simple boundaries.

It is important that these boundaries have no intersections, except at the parameter bounds, $\Delta=0$ or $\Delta=\infty$. Again we emphasize this is only valid for our particular curves $\hat{\bm \chi}$ and for spaces $\hat{Z}_k$ with consecutive $k$. As a counter-example, one can check that in the space $(\hat{Z}_0,\hat{Z}_2)$ the boundaries do have nontrivial self-intersections.

Now let us move on to ${\bf A}_3={\bf A}(0;1,1,1)$. Using the fact that the boundary of a Minkowski sum of objects is included in the Minkowski sum of the objects' boundaries, we simply compute the Minkowski sum of the boundaries of ${\bf A}_2$ and an extra curve. The resulting space is given in Figure \ref{F0b}, where the boundaries are ${\bf A}(0;3)$, ${\bf A}(0;1)$, ${\bf A}(1;1)$, and ${\bf A}(2;1)$. 

\begin{figure}[H]
  \centering
\includegraphics[height=2in]{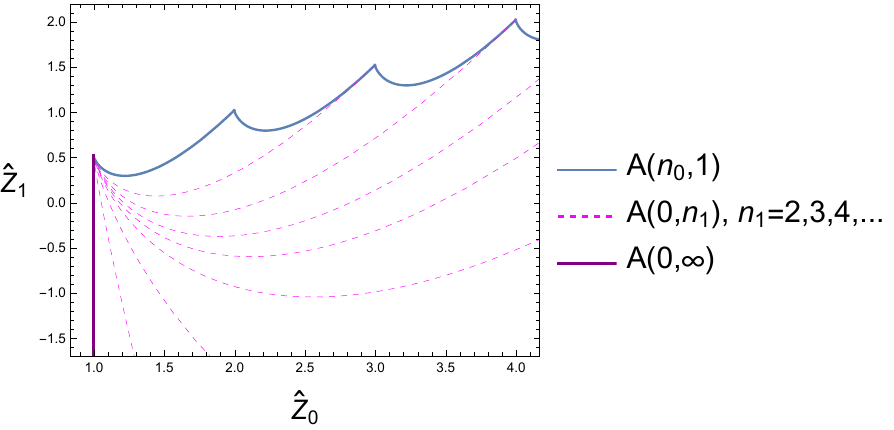} 
\caption{The lower boundary converges to a vertical line for $N\rightarrow \infty$}
  \label{F2a}
\end{figure}

This pattern persists as we consider $N$ curves, where we find the boundaries given by
\eqa\label{2dN}
\nonumber \textrm{upper bdy : } &{\bf A}(n_0,1)&, \ n_0=\{0,1,2,\ldots,N\}\,,\\
 \textrm{lower bdy : }& {\bf A}(0,N)&\,.
\eqae
This is shown in Figure \ref{F2a}, with the lower boundary being pushed successively downward. In the $N\gg1$ limit, the lower boundary (corresponding to ${\bf A}(0,\infty)$) becomes a straight vertical line. Meanwhile the upper boundary is simply the same periodic structure, repeated infinitely. We therefore obtain the boundaries of ${\bf A}_\infty$ in the 2D space $\{\hat{Z}_0,\hat{Z}_1\}$ as
\eqa\label{2dinf}
\nonumber \textrm{upper bdy : } &{\bf A}(n_0,1)&, \ n_0=\{0,1,2,\ldots\}\,,\\
 \textrm{lower bdy : }& {\bf A}(0,\infty)&\,.
\eqae

\begin{figure}[H]
  \centering
\includegraphics[height=2in]{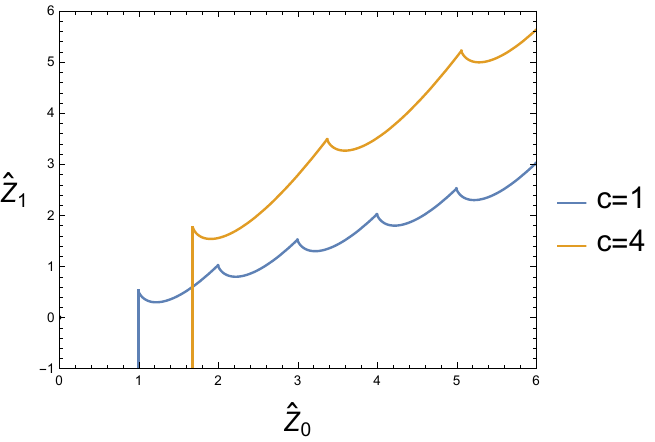} 
\caption{$N=\infty$ boundaries for $c=1$ and $c=2$.}
  \label{F2b}
\end{figure}

We also compare the $c=1$ and $c=2$ boundaries of this space in Figure \ref{F2b}.

\subsubsection{Three dimensions}
\begin{figure}[H] %  figure placement: here, top, bottom, or page
   \centering

   \begin{subfigure}[b]{0.48\textwidth}
         \centering
    \includegraphics[height=3.1in]{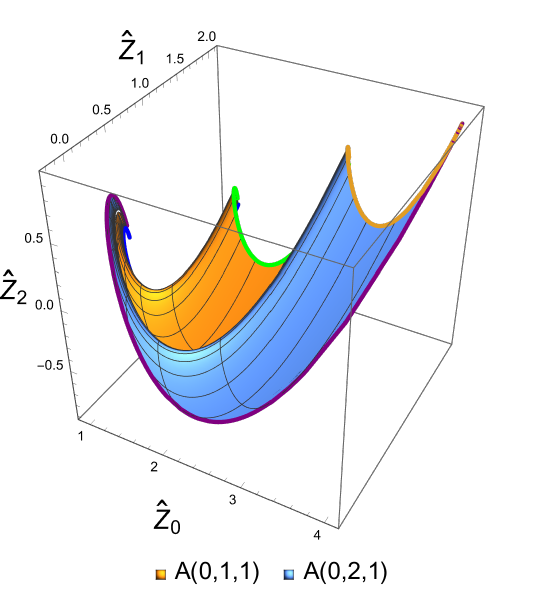}  
    \caption{Lower boundaries}
    \end{subfigure}
    \begin{subfigure}[b]{0.48\textwidth}
         \centering
     \includegraphics[height=3.1in]{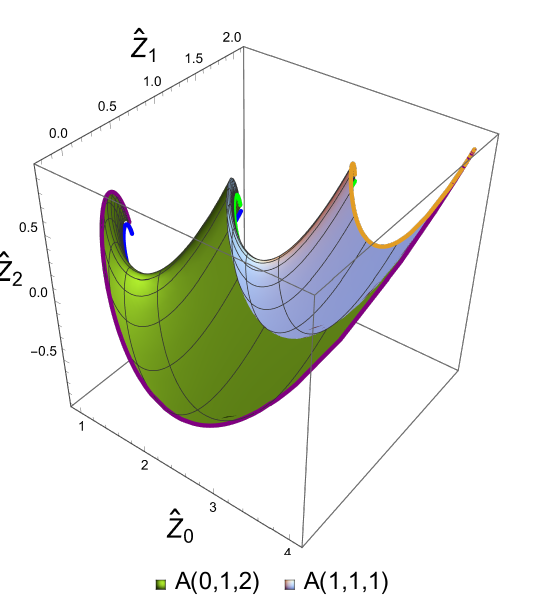} 
      \caption{Upper boundaries}
     \end{subfigure}
         \caption{Boundaries for the sum of three curves. The co-dimension 1 boundaries intersect along co-dimension 2 boundaries, which in turn are just the previous boundaries of the 2D space.}
   \label{F3}
\end{figure}

In 3D, for the space $(\hat{Z}_0, \hat{Z}_1, \hat{Z}_2)$, we begin with the Minkowski sum of three curves ${\bf A}_3\equiv {\bf A}(0;1,1,1)$, which corresponds to a 3D region, parameterized by the three $\m_i$. As before, the boundaries of this space can originate either from bounds of the parameters $\Delta=\{0,\infty\}$, or by extremizing via a Lagrange multiplier. We find the co-dimension one boundaries are given by ${\bf A}(0,2,1)$, ${\bf A}(0,1,2)$, ${\bf A}(1;1,1)$, and ${\bf A}(0;1,1)$,  and illustrate them in Figure \ref{F3}.
\noindent  

The generalization for $N$ curves can be obtained just as before by adding one extra curve at a time. We obtain the boundaries of ${\bf A}_N$ as
\eqa
\nonumber \textrm{lower bdy : } &{\bf A}(0,n_1,1)&, \quad n_1=\{1,2,\ldots,N{-}1\}\,,  \\
\textrm{upper bdy : } &{\bf A}(n_0,1,N{-}n_0)&, \quad n_0=\{0,1,2,\ldots,N{-}1\}\,.
\eqae
As a consistency check, it is easy to verify the lower and upper boundaries in 3D intersect along the co-dimension 2 boundaries given by eq. \ref{2dN}.

Taking $N\rightarrow \infty$ the upper boundary becomes ${\bf A}(n_0,1,\infty)$, which is a curved infinite strip in the $\hat{Z}_2$ direction for each $n_0$. The three components of the upper boundary can be parameterized as:
\eq
\begin{pmatrix}A^{(0)}(n_0,1,\infty)\\A^{(1)}(n_0,1,\infty)\\A^{(2)}(n_0,1,\infty)\end{pmatrix}=\begin{pmatrix}A^{(0)}(n_0,1)\\A^{(1)}(n_0,1)\\A^{(2)}(n_0,1)\end{pmatrix}+\begin{pmatrix}0\\0\\x\end{pmatrix}\,,
\eqe
with $x\ge0$ an independent parameter. This parameterization makes the relation between the geometry of $2D$ and $3D$ clear: projecting out $\hat{Z}_2$, we are left with $A(n_0,1)$, so the upper boundary from $3D$ becomes the lower boundary in $2D$. 

\begin{figure}[H] %  figure placement: here, top, bottom, or page
   \centering
\includegraphics[height=3.5in]{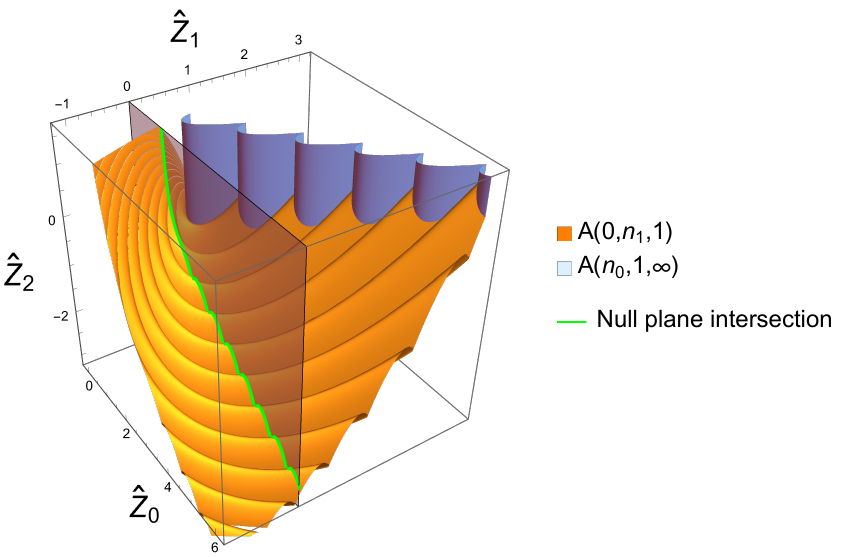} 
         \caption{Boundaries of ${\bf A}_\infty$ in three dimensions. $\hat{Z}_1=0$ is the first modular plane, intersecting the lower boundaries along the green line. This intersection can be worked out in closed form.}
   \label{F4}
\end{figure}

We therefore obtain the boundaries of ${\bf A}_\infty$ in 3D as
\eqa
\nonumber \textrm{lower bdy: }& {
\bf A}(0,n_1,1)\,, \quad n_1=\{1,2,3,\ldots\}\,,\\
\textrm{upper bdy: }& {\bf A}(n_0,1,\infty)\,, \quad n_0=\{0,1,2,\ldots\}\,.
\eqae
We illustrate these in Figure \ref{F4}, together with the null plane $\hat{Z}_1=0$. We will return to null planes shortly.

\subsubsection{Higher dimensions}
In 4D, the boundaries follow a similar pattern, and we conjecture this holds for arbitrary dimensions as well. We summarize our findings below, for a space $(\hat{Z}_0,\hat{Z}_1,\dots,\hat{Z}_{D-1})$
\begin{center}
\begin{tabular}{ |c|c|c|c|c|c| } 
 \hline
 & $D=2$ & $D=3$ & $D=4$& $D=5$ &\ldots \\ 
 \hline
 upper & ${\bf A}(n_0;1)$  & ${\bf A}(n_0,1,\infty)$& ${\bf A}(n_0,1,n_2,1)$ & ${\bf A}(n_0,1,n_2,1,\infty)$&\ldots\\ 
 lower &  ${\bf A}(0;\infty)$ & ${\bf A}(0,n_1,1) $ &   ${\bf A}(0,n_1,1,\infty)$ & ${\bf A}(0,n_1,1,n_2,1)$ & \\ 
 \hline
\end{tabular}
\end{center}
where $n_0=\{0,1,\ldots\}$, $n_i=\{1,2,\ldots\}$ for $i>0$, and ${\bf A}(n_0,n_1,\ldots)$ is given in eq.(\ref{Adef}), and parametrized by $\m_i\ge 0$. We next discuss the generalization when  $\m\ge \M$, and when a gap state is assumed to be present.

\subsection{Integrality with a gap}
The presence of a gap further reduces the space. There are two possible independent changes to be made. 

First, if we only assume that all states satisfy a bound $\Delta\ge \M$, the boundary structure remains unchanged, but the range of parameters is now $\Delta\in[\M,\infty)$. 
This does not automatically assume the state corresponding to $\M$ is actually present. Rather, the space contains all theories in which the lowest state is anywhere above this gap. 

Second, if we know that the state $\M$ itself is present, the space is modified further. This is only true for integer degeneracy, since if some state is present, it must have degeneracy equal to at least 1, and so must have a finite contribution to the partition function. Therefore, if we assume a gap state is present, all boundaries we found previously must have $n_0\ge1$. This amounts to translating the whole space by the vector $\hat{\bm {\chi}}(\M)$.

%Putting together these two conditions, namely the new range of parameters $\Delta\in[\M,\infty)$, and the translation by  $s(\M)$, we can carve out the allowed space for a theory with lowest state $\M$. We plot an example for $\M=1/10$ and $c=1$ for the space $(z_0,z_1)$.

\subsection{Null planes}
\begin{figure}[H] %  figure placement: here, top, bottom, or page
   \centering
\includegraphics[height=2.8in]{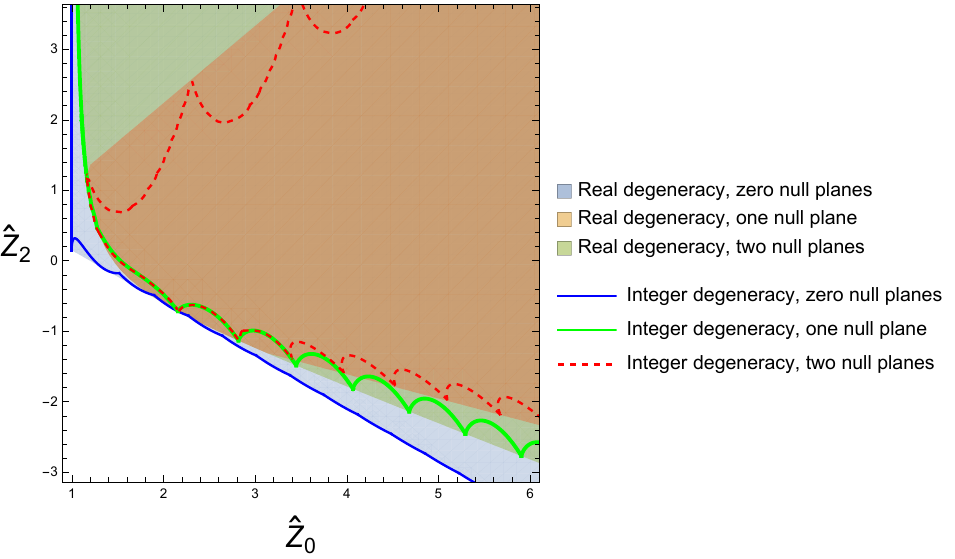} 
         \caption{Intersection of the geometry with 0, 1, and 2 null planes for $c=1$, comparing integer and non-integer degeneracy boundaries.}
   \label{F5}
\end{figure}

In this geometrical approach, modular invariance acts a series of null planes intersecting the geometry we described earlier. The intersection with the first null plane at $\hat{Z}_1=0$ was sketched in Figure \ref{F4}, and in fact can be solved analytically in terms of the Lambert $W_k(x)$, or product log function. This function is the solution to the equation
\eq
we^w=x\,,
\eqe
and has two real branches for $x\ge -e^{-1}$: the $k=0$ branch for $x\ge 0$, and the $k=-1$ branch for $x<0$. Intersecting a boundary of the form $A(0,n_1,1)$ with this null constraint implies solving
\eq
A^{(1)}(0,n_1,1)=G_{vac}^{(1)}+n_1 \hat{\chi}^{(1)}(\Delta_1)+n_2\hat{\chi}^{(1)}(\Delta_2)=0\,,
\eqe
giving 
\eq
\Delta_2= \frac{c-1}{12}+\frac{1}{4 \pi }-\frac{1}{2 \pi }W_k\left(\sqrt{e}\left( -G_{vac}^{(1)}- n_1 \hat{\chi}^{(1)}(\Delta_1)\right)\right),\quad k=0,-1\,,
\eqe
which at $c=1$ we find gives valid solutions for $n_1\ge 2$. This solution corresponds to the green intersection line in Figure \ref{F5}. For higher $c$, the first intersection occurs at higher $n_1$, with $n_1=7$ at $c=2$ for example. 
 
The intersection with higher order null planes can only be solved numerically. For instance, intersecting the 4D geometry upper boundaries with two null planes requires solving the intersection
\eq
A^{(1)}(n_0,1,n_2,1)=A^{(3)}(n_0,1,n_2,1)=0\,.
\eqe
We accomplish this in Mathematica by placing $\Delta_3$ on a grid, and using NSolve to find $\Delta_1$ and $\Delta_2$ for all values of $n_0$, $n_2$, and $\m_3$. Since high values of $\Delta$ are exponentially suppressed, we can optimize the search by limiting to low values. Experimentally we find $\Delta<10$ is sufficient. Once a valid solution for some values $(n_0,n_2)$ is found, the complete space can be quickly determined using the fact that consecutive values for $n_0,n_2$ correspond to neighboring boundaries.

We show the result for $c=1$ with zero, one, and two null plane intersections in Figure \ref{F5}, comparing with the case $n_i\in \mathbb{R}$ as well. We observe that the integrality condition rules out significant regions of allowed space, and only intersects the original boundary in isolated points. 

For even higher null constraints, the task becomes computationally intensive, and would require more specialized approaches that can efficiently find roots of polynomial-exponential equations. Note that we are not claiming the bounds due to integrality have converged at $\Lambda_{max}=3$. However, they are still valid bounds, as increasing the number of null plane intersections can only reduce the allowed space further.

%To add a third null plane we need to go to a 6D space $(a_0,a_1,\ldots,a_5)$. Assume the form is
%\eq
%A(n_0,n_1,n_2,1,1,1),\quad A(0,\infty,n_2,n_3,1,1)
%\eqe
%Then we have to solve a three null plane intersection of the above spaces, for three parameters. This would give surfaces in a space $(a_0,a_2,a_4)$, which can be further projected down.

\begin{figure}[H] %  figure placement: here, top, bottom, or page
   \centering
\includegraphics[height=2.8in]{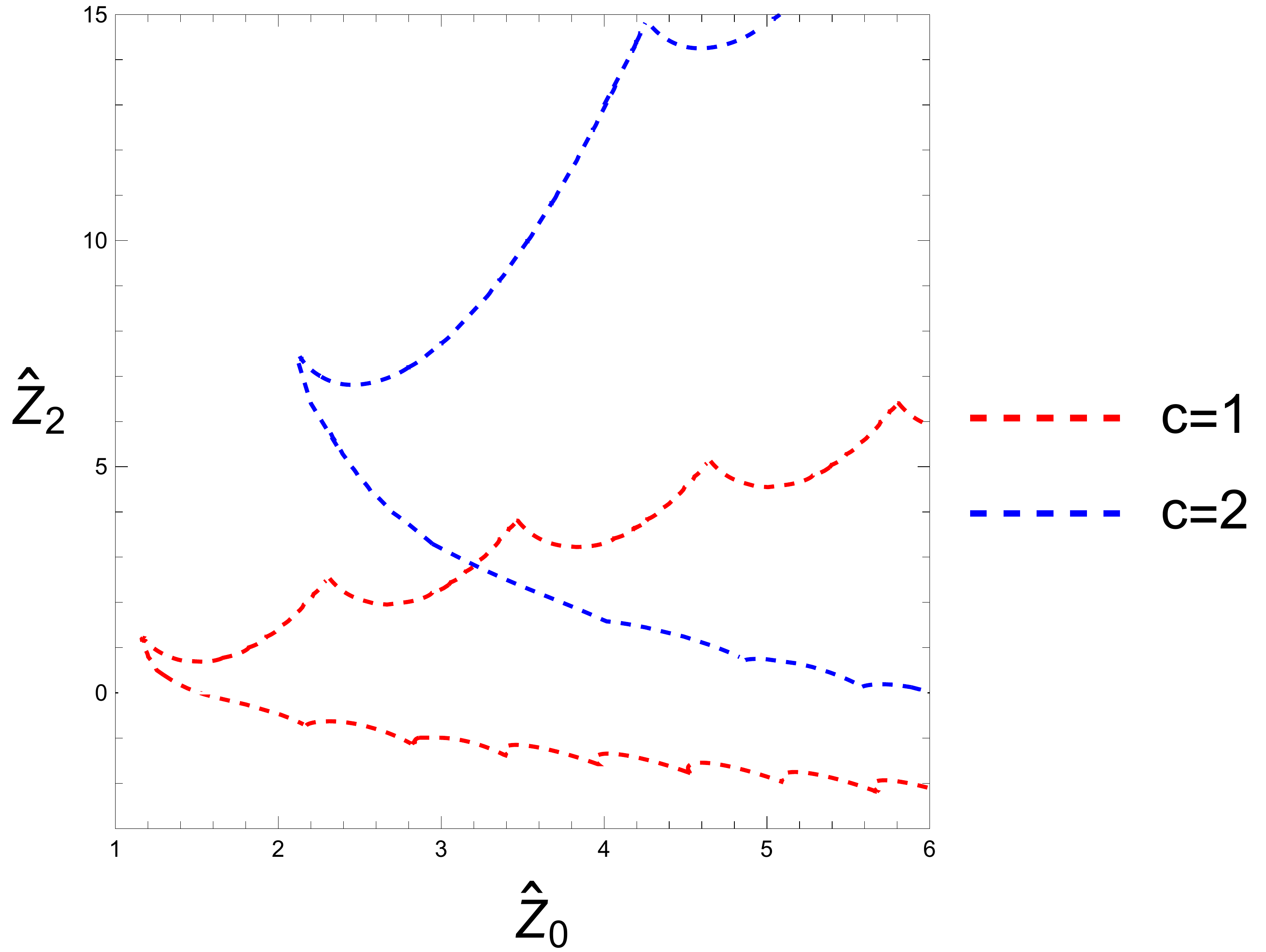} 
         \caption{Integer degeneracy boundaries for $c=1$ and $c=2$, at $\LL=3$.}
   \label{FC}
\end{figure}

\paragraph{Integrality at large $c$}
As we increase $c$, keeping $\LL=3$, we find that the relative importance of the integrality condition diminishes. In Figure \ref{FC} we compare the boundaries for $c=1$ and $c=2$.

However, it is known from the usual bootstrap approaches that as $c$ increases, we have to also increase $\LL$ for bounds to converge. Our current analysis is therefore not sufficient to make any claim on the effect of integrality at large $c$, but leaves open the possibility for significant effects at large also $c$.

\paragraph{Bounding the gap}
By increasing $\M$, eventually the null planes no longer intersect the geometry, allowing us to find bounds on the gap. With two null planes and $c=1$, we find a gap $\M\le 0.605 $, compared with $\M\le 0.615$ from the usual spinless bootstrap with no restriction on degeneracy, when using the same number of null planes. Using more null planes, the spinless bootstrap gap bound converges to  $\M\le 0.603$, while the spinning bootstrap converges to $\M\le 0.5$, saturating the free boson theory. Since the bound is already saturated by the spinning bootstrap there is no improvement to be found, however we find it encouraging that the integer condition is stronger than the spinless bootstrap at equivalent derivative order. This leaves open the possibility that at large $\Lambda_{\textrm{max}}$ the integer condition improves the spinless result, and for large $c$ it may improve also the spinning bootstrap bound, given the reductions in allowed theory space we found previously.

\subsection{Comparing with non-integer bootstraps and known theories at $c=1$}
We can now compare the integer bootstrap at $\Lambda_\textrm{max}=3$ with the SDP bootstrap at much higher orders, and also to the known $c=1$ free boson theory.

For $c=1$, the character representation in eq.(\ref{Blocks}) is no longer valid. This is because  for $c=1$, primaries with $h=n^2$ and $h=(n+\frac{1}{2})^2$, $n\in\mathbb{Z}$ have extra null states that requires removal, introducing minus signs in the usual character expansion. Therefore, in general for $c=1$, our geometry requires modification to include these special primaries. 

However, for compactified free bosons at special  radius its partition function can be positively expanded on the usual characters. Indeed its reduced partition function is given by \cite{Ginsparg:1987eb}
\eq
Z(R)=\left(\tau \bar{\tau}\right)^\frac{1}{4}\sum_{a,b=-\infty}^\infty q^{(a/R+b R)^2/4}\bar{q}^{(a/R-b R)^2/4}\,,
\eqe
containing states 
\eq\label{circlestate}
(h,\bar{h})=\left(\frac{1}{4}\left(\frac{a}{R}+b R\right)^2,\frac{1}{4}\left(\frac{a}{R}- b R\right)^2\right)\,,
\eqe
with $a,b\in \mathbb{Z}$. To express this partition function in the form of eq.(\ref{Blocks}), we need to subtract the vacuum $(1-q)(1-\bar{q})$, and for general $R$ the term $q\bar{q}$, corresponding to a state $(h,\bar{h})=(1,1)$, appears with a minus sign. However, for $R=\frac{n}{2}$, $n\ge 2$, $n\in \mathbb{Z}$, the term becomes positive. Therefore, it becomes a valid question if these specific theories can be bootstrapped. There are three interesting cases we can consider for the bootstrap, depending on our assumptions of the gap state.

\begin{figure}[H] %  figure placement: here, top, bottom, or page
   \centering       
            \includegraphics[width=0.8\textwidth]{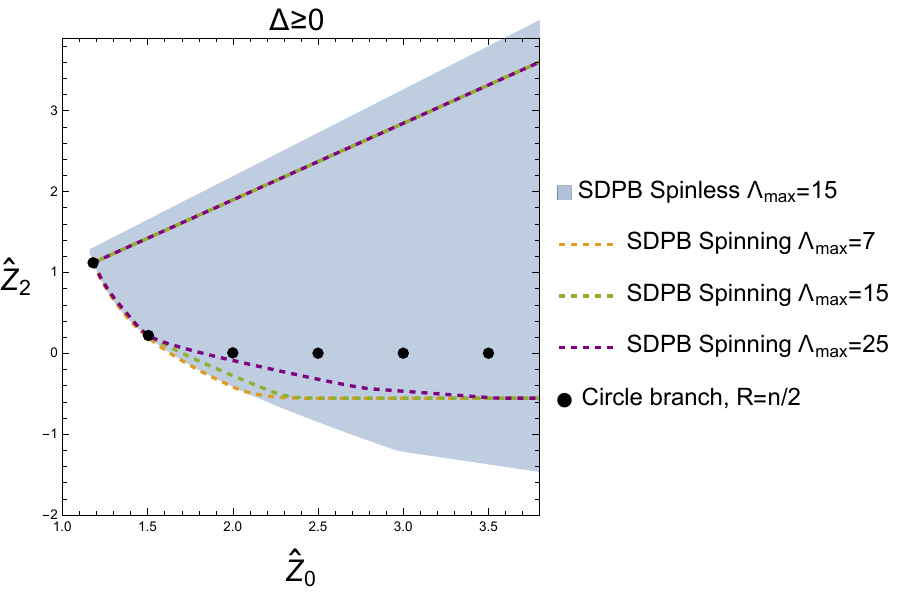}
            \caption{Regions carved out by SDPB, for $c=1$, with no assumption on the scaling dimension except $\Delta\ge 0$. The black dots represent the circle branch of the free boson theory, evaluated at particular $R=\frac{n}{2}$, $n\in\mathbb{N}$. The spinless bootstrap has already converged around $\LL=15$, while the spinning bootstrap region is still shrinking. At large $\LL$ we conjecture the lower boundary will reach the circle branch theories. }
   \label{G1}
\end{figure}

\paragraph{$\m\ge 0$, no gap state assumed}
We compare the case where the only assumption on the scaling dimensions is $\m\ge 0$, but we do not assume a state with $\m=0$ is necessarily present. In other words, the allowed region includes all possible theories with any gap. This is shown in Figure \ref{G1}. 
First, we explore convergence from usual SDPB without imposing integrality. We find that spinless converges quickly around $\LL=15$, while spinning has still not converged in the bottom region even at $\LL=25$. It appears plausible that this lower boundary will converge precisely on the circle branch theories.

Next we plot the boundary due to imposing the integrality (up to $\LL=3$) on the spinless bootstrap. We observe this boundary is stronger in many regions compared to both the spinless and spinning bootstrap at much higher order. Since the circle branch points can only be saturated by non-convex conditions, it is possible that integrality at high $\LL$ may converge exactly on these isolated points.

 \begin{figure}[H] %  figure placement: here, top, bottom, or page
   \centering       
            \includegraphics[width=0.8\textwidth]{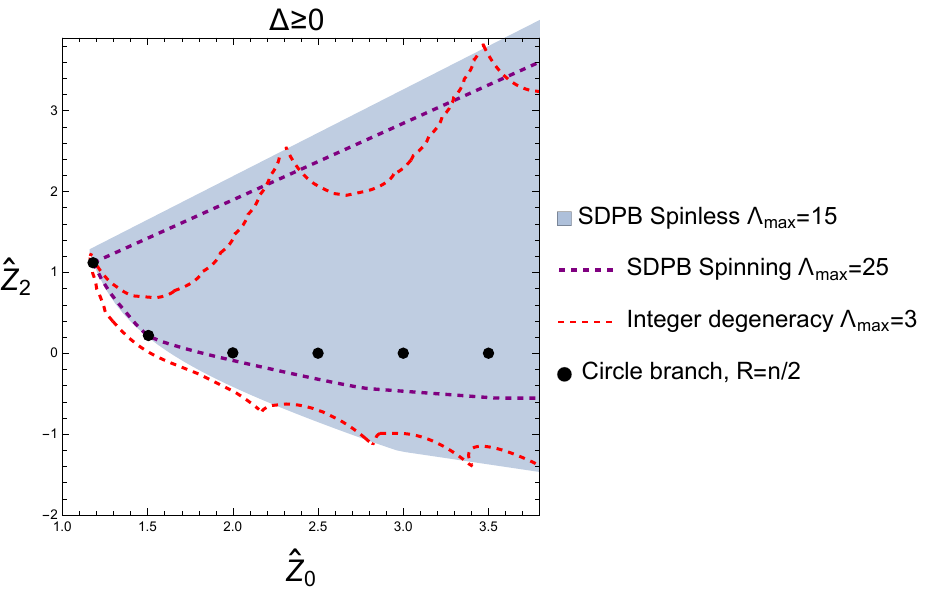}
            \caption{Comparing regions carved out by SDPB with the region carved out by the (spinless) degeneracy integrality condition at $\LL=3$. The integrality condition significantly approaches the circle branch theories.}
   \label{G1b}
\end{figure}

\paragraph{$\Delta\ge  0.1$, gap state not present}
It is also instructive to explore this space when we require all $\Delta\ge0.1$ as used for Figure \ref{fig:z0z2gap}. Here we are still not assuming the presence of any particular gap state, only that it must be somewhere above 0.1. In the circle branch, the lowest state in scaling dimension corresponds to $a=1,b=0$, implying that the circle branch satisfies $\Delta\ge \M$ if $R\le (4 \M)^{-\frac{1}{2}}$, leading to three allowed theories at $R=\{1,3/2,2\}$ for this particular gap. 

\begin{figure}[H] %  figure placement: here, top, bottom, or page
      \centering
                 \includegraphics[width=0.8\textwidth]{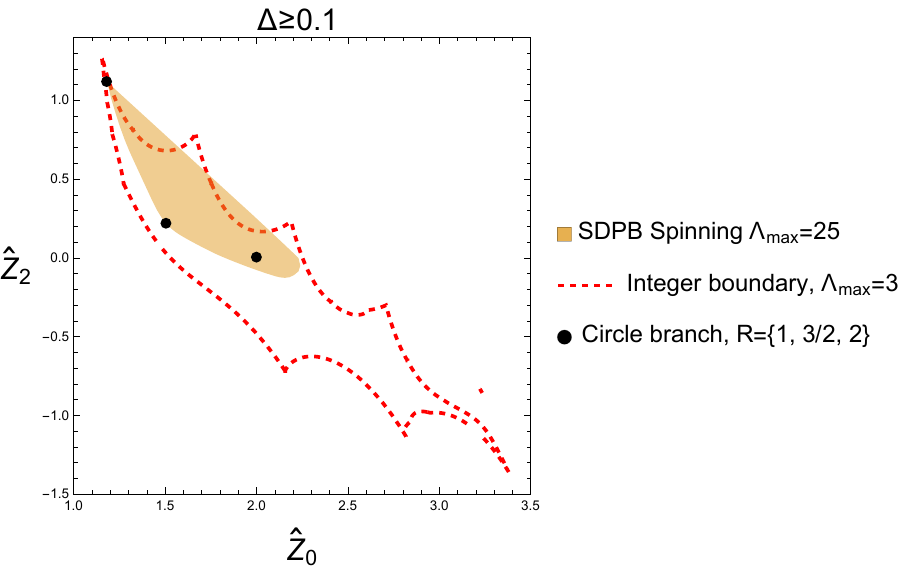}
                 \caption{States satisfy $\Delta\ge 0.1$, but gap state is not necessarily at $\Delta=0.1$. Three circle branch theories are compatible with this restriction.\\}
                 \label{F10A}
                 \end{figure}

\begin{figure}[h] %  figure placement: here, top, bottom, or page
      \centering
                    \includegraphics[width=0.8\textwidth]{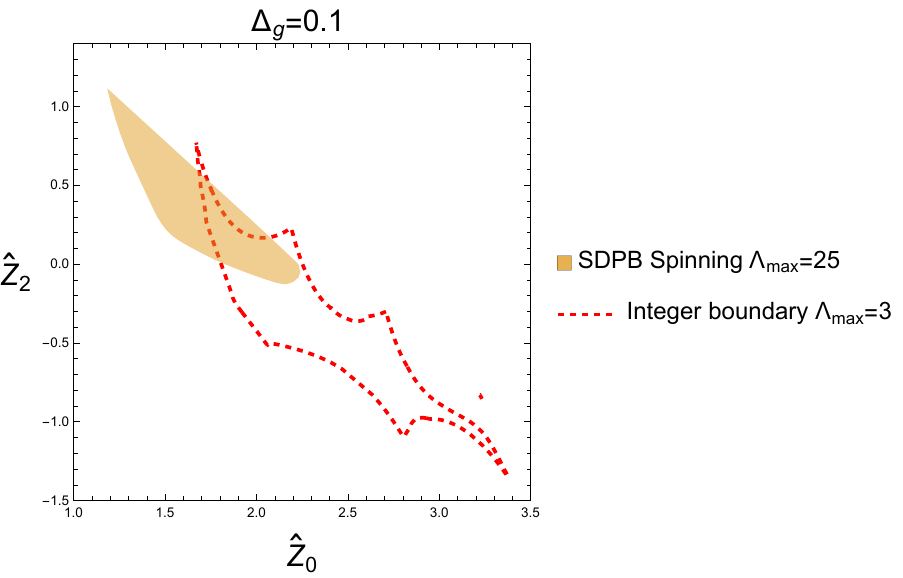}
                     \caption{Gap state has $\M= 0.1$, and is assumed to be present, significantly pushing the integer boundary. There are no know theories satisfying our constraints for this gap.}
                      \label{F10B}                
   \label{F10}
\end{figure}

\paragraph{$\M=  0.1$, gap state present}
In Figure \ref{F10B} we plot the allowed region of theories which have the gap state at $\M=0.1$. For this particular gap these is no known theory to exist.  We see that the integer condition rules out an even larger piece of the region allowed by the spinning bootstrap.

\begin{figure}[h]
  \centering
  \includegraphics[width=0.8\textwidth]{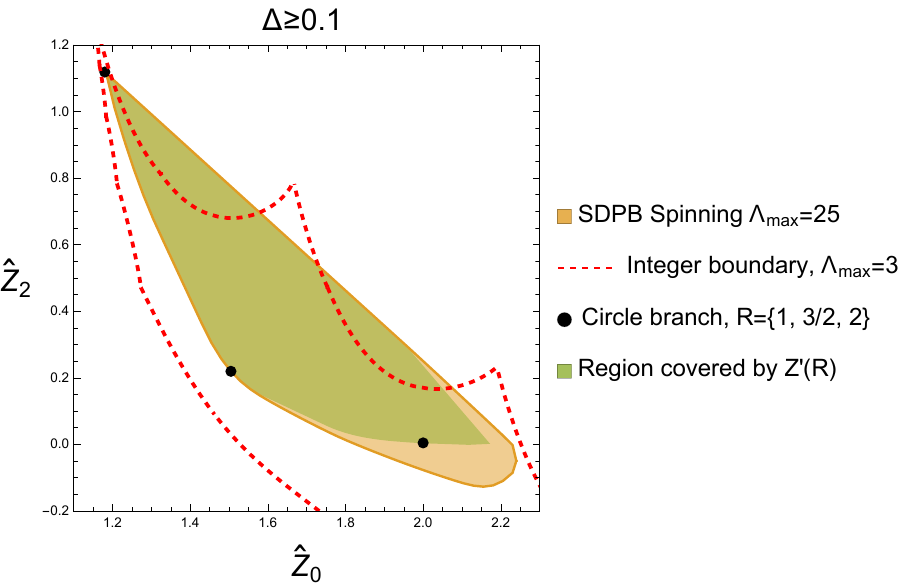}
  \caption{The space covered by eq.(\ref{linearcomb}) covers most of the space allowed by the spinning bootstrap. The integrality condition starts ruling out some of this space.}
  \label{compare}
\end{figure}

\paragraph{Linear combinations of circle branches}
As observed in \cite{Collier:2016cls}, it is possible to take linear combinations of circle branch theories which have positive expansion on non-degenerate characters for any $R$. Let us focus on the space for $\Delta\ge 0.1$, where the only allowed partition functions compatible with our assumptions are for $R=\{1,3/2,2\}$, and consider the combination
\eq\label{linearcomb}
Z'(R)=(1-x-y-z)Z(R)+xZ(1)+yZ(3/2)+z Z(2)\,.
\eqe
It is easy to verify that for particular ranges of $x,y,z$ the above partition function has positive expansion on the non-degenerate characters, as well as the correct vacuum normalization. Clearly, for generic values of $R$ it will not satisfy integer degeneracy, and so does not constitute a valid CFT, but it will be compatible with the spinning bootstrap. We plot this in Figure \ref{compare}, and note this simple linear combination covers most of the space except a corner, at least for $\LL=25$. This is similar to what was observed in the EFT bootstrap, where a linear combination of two theories spanned most of allowed space \cite{Caron-Huot:2020cmc}. We can also view this result in the other direction: the bootstrap, based only on convex conditions like positivity, will never be able to rule out the region covered by this linear combination. Adding the integrality condition is therefore necessary to reduce this space further, and we can see that this occurs already at $\LL=3$.

\iffalse
\subsection{Spinning sneak peak}
If we consider the full geometry, we need to compute spaces of the type (after some simplification)
\eq\label{spinint}
a_{k,q}=\sum_{i}n_i J_i^q e^{-\Delta_i}\Delta_i^k 
\eqe
This can still be viewed as a Minkowski sum, and as before we can instead simply consider the sum
\eq
a_{k,q}=\sum_{i}J_i^q e^{-\Delta_i}\Delta_i^k 
\eqe
since by picking $n_i$ state to have equal $J$ and $\Delta$ we can reproduce any term in eq. \ref{spinint}. We now need to compute the Minkowski sum of non-identical curves.   The approach was described in Ref., but 
\fi

%%%%%%%%%%%%%%%%%%%%%%%%%%%%%%%%%%%%%%%%%%%%%%%%%%%%%
\section{Conclusions}\label{section6}
%%%%%%%%%%%%%%%%%%%%%%%%%%%%%%%%%%%%%%%%%%%%%%%%%%%%
In this paper, we elucidate the geometry behind the modular bootstrap, where the constraint of modular invariance is imposed on the derivative expansion of the torus partition function around the self dual point. We show that the character representation of the torus partition function can be mapped to the convex hull of a double moment curve on $\mathbb{R}^+\otimes \mathbb{Z}$, while modular invariance acts as a series of planes intersecting this space. 

 The spinless bootstrap corresponds to a simpler single moment problem, where sufficient conditions are well known. For the double moment problem which includes the spin related constraints, we only have a collection of necessary conditions. In both cases, these conditions are phrased in terms of the total positivity of (generalized) Hankel matrices constructed from the Taylor coefficients of the partition function. This allows us to efficiently identify the boundary analytically at low derivative order, but also at much higher order via numeric methods such as semi-definite programming. 

We extract the optimal gap and twist gap for a wide range of values for the central charge, at derivative order $\LL$ up to 43. This is not sufficient to match results from the highly efficient fast bootstrap method. However, one advantage of our approach is that the value of central charge does not affect the computation time, allowing us to directly obtain bounds above $c\gtrsim 2000$. Furthermore, while the bounds we derive depend on spin truncation, they are always valid bounds. This is because by our construction raising the spin truncation can only tighten the bounds. As such one can eliminate the step of checking convergence with spin truncation, which is necessary in traditional SDPB approaches.

Next, we observe that the allowed region of Taylor coefficients is in general unbounded unless the spectrum has a sufficiently large gap. This allows us to introduce the notion of a critical gap $\Delta^*_{\text{gap}}$ above which $\hat{Z}_0$ develops an upper bound and the partition function is bounded from all directions. Such upper bound can also be extracted from the HKS~ \cite{Hartman:2014oaa,Mukhametzhanov:2019pzy} bound, evaluated at $\tau=i$, leading to $\Delta^*_{\text{gap}}=\frac{c{-}1}{12}{+}\frac{1}{4\pi}$. We demonstrate that the ``optimal" critical gap is in fact lower than this value, and that it tends to $\frac{c}{12}$ in the large $c$ limit. Furthermore, in the latter limit we observe multiple kinks for $\Delta^*_{\text{gap}}\leq\Delta_{\text{gap}}\leq\frac{1}{12}$ that are stable against increase in derivative expansion. It will be interesting to understand the nature of these kinks and whether they correspond to physical theories. For pure quantum gravity the gap is at $\frac{c{-}1}{12}$, thus the space is unbounded and we only have a lower bound for $\hat{Z}_0$. Preliminary results indicate that: 
\begin{align}
 \hat{Z}_0 \gtrsim 1.180 
\end{align}
with mild dependence on $c$. This can be compared to the newly proposed partition functions in~\cite{DiUbaldo:2023hkc}, which are conjectured to be unitary at particular values of $c$. We leave the detailed analysis of such lower bounds for future work.

We then compare the allowed space obtained via the bootstrap to known theories at $c=1$. We find that even though spinning bootstrap does not improve the gap bound, it still drastically reduces allowed space, with the free boson theories living partially on the boundary of this space. Curiously, linear combinations of these theories in fact almost saturate allowed space. These combinations are taken such that they respect positivity and vacuum normalization, but generically they do not satisfy integer degeneracy, and so are not consistent CFTs. 

This motivates us to explore the consequences of integrality, and we indeed find it can be used to rule out parts of allowed space, even starting at very low derivative order. We show that this integrality condition can be imposed by computing a Minkowski sum of identical curves, which leads to a non-convex geometry. Intersecting this geometry with up to two modular planes ($\LL=3$) we find that integrality rules out regions allowed by the spinning bootstrap at much higher $\LL$. We speculate that at high $\Lambda_{\textrm{max}}$, integrality could in fact reduce the space to precisely the free boson theories. On bounding the gap, we find integrality is stronger than the spinless bootstrap at equivalent derivative order, but at high $\Lambda$ or high $c$ it is still not clear what the effect of integrality is. The reductions in theory space support the possibility that integrality should also improve bounds on the gap, but direct computation is required for a definitive conclusion.

There are several questions we have not yet answered about integrality. First would be pushing the derivative order, ie. the number of null plane intersections. Since the boundaries are known to arbitrary dimension, this problem is purely computational, where one has to solve systems of polynomial-exponential equations. More efficient and specialized algorithms could potentially increase $\Lambda_{\textrm{max}}$, hopefully to the point where bounds begin to converge and become relevant also at large $c$. Finally, we have not explored the combination of imposing both integer spin and integer degeneracy. Our approach of finding boundaries via Minkowski sums can still be used in principle, but organising the geometry in such a high dimensional space becomes a very difficult task. Solving both integer spin and integer degeneracy to high order in $\Lambda_{\textrm{max}}$ would finally answer if the modular bootstrap alone can fix $\M=\frac{c}{12}$, as expected from holography, and more generally how the space of known theories compares to the space of allowed theories.

It should also be feasible to apply our approach to superconformal characters. For N=2 super conformal symmetry, the states are now labeled by $(h, \bar{h}, Q, \bar{Q})$, where $Q$ is the charge under the U(1) R symmetry, and the partition function is given by sums over BPS and non-BPS representations (see see~\cite{Keller:2012mr, Friedan:2013cba} for details). For the spinless bootstrap, the space is given by four distinct moment curves, subject to modular constraints that interwine the four curves. It will be interesting to see if the space of consistent partition function also exhibits kink like behaviour with some critical gap.

%%%%%%%%%%%%%%%%%%%%%%%%%%%%%%%%%%%%%%%%%%%%%%%%%%%%%
\section*{Acknowledgements}
%%%%%%%%%%%%%%%%%%%%%%%%%%%%%%%%%%%%%%%%%%%%%%%%%%%%
We thank Scott Collier, Shu-Heng Shao and Yuan Xin for discussions. LYC, YTH and HCW are supported by MoST Grant No. 109-2112-M-002 -020 -MY3 and 112-2811-M-002 -054 -MY2. WL is supported by the US Department of Energy Office of Science under Award Number DE-SC0015845, and the Simons Collaboration on the Non-Perturbative Bootstrap. T.-C. H. is supported by the Simons Collaboration on Global Categorical Symmetries. This work was performed in part at the Aspen Center for
Physics, which is supported by National Science Foundation grant PHY-2210452. This project has received funding from the European Union’s Horizon 2020 research and innovation programme under the Marie Skłodowska-Curie grant agreement No 101025095. 
\appendix
\section{Flat extensions of the single moment problem}\label{App:flat}
Having positive semi-definite Hankel and shifted-Hankel matrices are not sufficient constraints for the single moment problem. This is best illustrated with an example. Consider $\vec{y}=(2,\sqrt{2},1,1)$, which gives the following Hankel and shifted Hankel matrix:
\eq\label{eq:Exp}
 K_1[\vec{y}]=\left(\begin{array}{cc}2 & \sqrt{2} \\ \sqrt{2} & 1\end{array}\right)\,,\quad K^{\rm shift}_1[\vec{y}]=\left(\begin{array}{cc}\sqrt{2} & 1 \\ 1 & 1\end{array}\right)\,,
\eqe
both satisfying eq.(\ref{ConvHalf}). However $\vec{y}$ is not in the hull since ${\rm Det}\, K_1[\vec{y}]=0$, implying there is only one element in the hull. But then ${\rm Det}\, K_1$ must then vanish as well, leading to a contradiction. 

This example provides a hint towards the correct sufficient conditions. First of all, we see that contradictions can only occur when minors vanish, implying a bound on the rank. Then Hankel and shifted-Hankel being positive definite are sufficient conditions for a solution to the moment problem. If some minors vanish, then one needs to further require that the extended (larger) minors also vanish. Indeed as discussed by Curto and Fialkow~\cite{Curto}, the sufficient condition for singular Hankel matrices is the requirement that there exits a \textit{flat extension} of  $K_n$ to  $K_{n{+}1}$, such that the latter is totally non-negative \textbf{and} has the same rank as $K_n$.

Starting with a singular $K_n$, write  
\[
K_{n{+}1}=\left(\begin{array}{@{}c|c@{}}
  K_n
  & b \\
\hline
  b^T &
  c
\end{array}\right),
\]
For $K_{n{+}1}$ to be positive semi-definite, we must have $v^T K_{n{+}1} v\geq 0$ for all $v$. Let's write $v=\left(\begin{array}{@{}c}
  x \\
\hline
  y
\end{array}\right)$. Then we have
\[
 x^T K_n x+2(x\cdot b)y+c y^2\geq 0.
\]
If $x$ lies in the null space of $K_n$, we then have $c y^2\geq 0$
\[
 c (y+\frac{x\cdot b}{c})^2-\frac{(x\cdot b)^2}{c}\geq 0.
\]
For the above to be always positive for all $y$, $b$ must be orthogonal to all vectors in the null space of $K_n$, which means $b$ shall be in the range of $K_n$. Furthermore, with $b$ being spanned by the vectors $K_n$, $K_{n{+}1}$ must be singular as it must. 

In particular, the sequence $(2,\sqrt{2},1, 1)$ fails to satisfy this requirement since $(1,1)\notin \text{Ran}(K_2)=\{b(\sqrt{2},1)|b\in \mathbb{R}\}$.

\section{Explicit details for SDP}
\label{appendix_SDP}
\paragraph{Standard form} Any semi-definite program can be brought into the standard form
\begin{align}
  \label{SDP_standard_form}
  \text{Min } & \quad \mathbf{c}^T \mathbf{x}  \\
  \label{SDP_constraints_1}
  \text{subject to } & \quad \sum_{i = 1}^{m} x_i A_i^{(l)} - C^{(l)} \succeq 0, \quad l = 1, ..., L, \\
  \label{SDP_constraints_2}
  & \quad B^T \mathbf{x} = \mathbf{b}, 
\end{align}
where
\begin{equation}
  \mathbf{x} \in \mathbb{R}^m, \quad A_i^{(l)} \in \mathbb{S}^{k^{(l)}}, \quad B \in \mathbf{R}^{m \times n}, \quad b \in \mathbf{R}^{n}.
\end{equation}
\paragraph{Carving out the theory space} Seen in section \ref{the_modular-hedron}, the problem of bounding the partition function can be viewed as a semi-definite program, where the optimization variable $\mathbf{x}$ consists of the discretized partition functions, i.e. the vectorization of (\ref{DiscreteZ}). In this basis the equality constraint (\ref{SDP_constraints_2}), which corresponds to the modular plane, becomes trivial. We simply remove the vanishing $z^{(i,j)}$ with odd $i+j$ and reduce the dimension of the problem by roughly half. Therefore, we almost never need to work with equality constraints explicitly. \(A_i^{(l)}\) and $C^{(l)}$ are symmetric matrices that depend on the gap and the central charge, and $L$ counts the number of positivity constraints, which is equal to the number of generalized Hankel matrices. Solving (\ref{SDP_standard_form}) then enables us to bound arbitrary linear combinations of the discretized partition functions. 
\paragraph{Bounding the gap} To find the maximal possible value of, say, the gap of the scaling dimension $\Delta_{\text{gap}}$, we start from a small value of $\Delta_{\text{gap}}$ and gradually increase its value until no solution $\mathbf{x}$ that satisfies the constraints (\ref{SDP_constraints_1}) can be found. Therefore, to obtain an upper bound for $\Delta_{\text{gap}}$, one needs a certificate of \emph{infeasibility}. To achieve this, we can introduce an auxiliary variable $t$ which measures the violation of the positivity, and solve the following optimization problem:
\begin{align}
  \text{Min} & \quad t \\
  \text{s.t.} & \quad \sum_{i = 1}^{m} x_i A_i^{(l)} - C^{(l)} + t I \succeq 0, \quad l = 1, ..., L.
\end{align}
Suppose the optimal value $t^*$ is positive, the problem is infeasible; otherwise, the problem is feasible. 
\paragraph{Certifying unboundedness} Suppose we obtain an optimal solution to (\ref{SDP_standard_form}), it automatically serves as a certificate of boundedness of $\mathbf{c}^T \mathbf{x}$. However, the optimization problem (\ref{SDP_standard_form}) may also be unbounded, such as when the gap $\Delta_{\text{gap}}$ is tuned to some value below the critical gap. The following feasibility problem can then be served as a certificate for unboundedness:
\begin{align}
  \text{find } & \quad \mathbf{y} \\
  \text{s.t. } & \quad \mathbf{c}^T \mathbf{y} < 0, \\
  & \quad \sum_{i = 1}^{m} y_i A_i^{(l)} \succeq 0. 
\end{align}
Since if such a point $\mathbf{y}$ exists, then $\textbf{x} + \alpha \textbf{y}$ for any $\alpha > 0$ will again be a feasible point of problem (\ref{SDP_standard_form}), implying that the objective function is unbounded from below. 
\paragraph{Implementation} There are many freely available solvers for semi-definite programs, for example \texttt{SDPA} \cite{nakata_2010} and \texttt{SDPB} \cite{simmons-duffin_2015}. Designed especially for polynomial matrix programs, the \texttt{SDPB} solver is widely used in the bootstrap community. Ordinary semi-definite programs are a special case of polynomial programs, which we frequently encounter in this paper. While \texttt{SDPB} is highly efficient and optimized for polynomial programs, it is overkill for solving ordinary SDPs. Instead, we needed a solver that could handle not only large number of spins but also large matrix sizes. To this end, we developed \texttt{SDPJ}, a native Julia SDP solver based on the primal-dual interior point method, which is parallelized and supports arbitrary precision, with a modified parallelization architecture to suit our needs. The code is publicly available on GitHub \cite{fishbonechiang}. The choice of parameters in this paper is shown in Table \ref{parameters}. 
\begin{table}[h]
  \center
  \begin{tabular}{|l|l|}
    \hline
    \texttt{beta}             &   $0.01$                  \\
    \texttt{Omega\_p}         &   $10^{10} \sim 10^{50}$  \\
    \texttt{Omega\_d}         &   $10^{10} \sim 10^{50}$  \\
    \texttt{epsilon\_gap}     &   $10^{-10} \sim 10^{-3}$ \\
    \texttt{epsilon\_primal}  &   $10^{-50}$              \\
    \texttt{epsilon\_dual}    &   $10^{-50}$              \\
    \texttt{prec}             &   $300$                   \\
    \hline
  \end{tabular}
  \caption{Choice of parameter for \texttt{SDPJ} in this work. See \cite{fishbonechiang} for the definition of the parameters.}
  \label{parameters}
  \end{table}

% \vspace{-0.4cm}

% \vspace{-0.3cm}

% \vskip .3 cm

\bibliography{refs2}

\providecommand{\href}[2]{#2}\begingroup\raggedright\begin{thebibliography}{10}

\bibitem{Arkani-Hamed2014-wv}
N.~Arkani-Hamed and J.~Trnka, \emph{The amplituhedron}, {\emph{J. High Energy
  Phys.} {\bfseries 2014} (2014) 30}.

\bibitem{Arkani-Hamed2018-nd}
N.~Arkani-Hamed, Y.~Bai, S.~He and G.~Yan, \emph{Scattering forms and the
  positive geometry of kinematics, color and the worldsheet}, {\emph{J. High
  Energy Phys.} {\bfseries 2018} (2018) 96}.

\bibitem{He:2022cup}
S.~He, C.-K.~Kuo, Z.~Li and Y.-Q.~Zhang, \emph{{All-Loop Four-Point
  Aharony-Bergman-Jafferis-Maldacena Amplitudes from Dimensional Reduction of
  the Amplituhedron}},
  \href{https://doi.org/10.1103/PhysRevLett.129.221604}{\emph{Phys. Rev. Lett.}
  {\bfseries 129} (2022) 221604}
  [\href{https://arxiv.org/abs/2204.08297}{{\ttfamily 2204.08297}}].

\bibitem{He:2023rou}
S.~He, Y.-t.~Huang and C.-K.~Kuo, \emph{{The ABJM Amplituhedron}},
  \href{https://arxiv.org/abs/2306.00951}{{\ttfamily 2306.00951}}.

\bibitem{Arkani-Hamed2017-qj}
N.~Arkani-Hamed, P.~Benincasa and A.~Postnikov, \emph{Cosmological polytopes
  and the wavefunction of the universe},
  \href{https://arxiv.org/abs/1709.02813}{{\ttfamily 1709.02813}}.

\bibitem{Adams:2006sv}
A.~Adams, N.~Arkani-Hamed, S.~Dubovsky, A.~Nicolis and R.~Rattazzi,
  \emph{{Causality, analyticity and an IR obstruction to UV completion}},
  \href{https://doi.org/10.1088/1126-6708/2006/10/014}{\emph{JHEP} {\bfseries
  10} (2006) 014} [\href{https://arxiv.org/abs/hep-th/0602178}{{\ttfamily
  hep-th/0602178}}].

\bibitem{Arkani-Hamed2021-pa}
N.~Arkani-Hamed, T.-C.~Huang and Y.-T.~Huang, \emph{The {EFT-hedron}},
  {\emph{J. High Energy Phys.} {\bfseries 2021} (2021) 259}.

\bibitem{deRham:2017avq}
C.~de~Rham, S.~Melville, A.J.~Tolley and S.-Y.~Zhou, \emph{{Positivity bounds
  for scalar field theories}},
  \href{https://doi.org/10.1103/PhysRevD.96.081702}{\emph{Phys. Rev. D}
  {\bfseries 96} (2017) 081702}
  [\href{https://arxiv.org/abs/1702.06134}{{\ttfamily 1702.06134}}].

\bibitem{Caron-Huot:2020cmc}
S.~Caron-Huot and V.~Van~Duong, \emph{{Extremal Effective Field Theories}},
  \href{https://doi.org/10.1007/JHEP05(2021)280}{\emph{JHEP} {\bfseries 05}
  (2021) 280} [\href{https://arxiv.org/abs/2011.02957}{{\ttfamily
  2011.02957}}].

\bibitem{Tolley:2020gtv}
A.J.~Tolley, Z.-Y.~Wang and S.-Y.~Zhou, \emph{{New positivity bounds from full
  crossing symmetry}},
  \href{https://doi.org/10.1007/JHEP05(2021)255}{\emph{JHEP} {\bfseries 05}
  (2021) 255} [\href{https://arxiv.org/abs/2011.02400}{{\ttfamily
  2011.02400}}].

\bibitem{Bellazzini:2020cot}
B.~Bellazzini, J.~Elias~Mir\'o, R.~Rattazzi, M.~Riembau and F.~Riva,
  \emph{{Positive moments for scattering amplitudes}},
  \href{https://doi.org/10.1103/PhysRevD.104.036006}{\emph{Phys. Rev. D}
  {\bfseries 104} (2021) 036006}
  [\href{https://arxiv.org/abs/2011.00037}{{\ttfamily 2011.00037}}].

\bibitem{Haldar:2021rri}
P.~Haldar, A.~Sinha and A.~Zahed, \emph{{Quantum field theory and the
  Bieberbach conjecture}},
  \href{https://doi.org/10.21468/SciPostPhys.11.1.002}{\emph{SciPost Phys.}
  {\bfseries 11} (2021) 002}
  [\href{https://arxiv.org/abs/2103.12108}{{\ttfamily 2103.12108}}].

\bibitem{Chiang:2021ziz}
L.-Y.~Chiang, Y.-t.~Huang, W.~Li, L.~Rodina and H.-C.~Weng, \emph{{Into the
  EFThedron and UV constraints from IR consistency}},
  \href{https://doi.org/10.1007/JHEP03(2022)063}{\emph{JHEP} {\bfseries 03}
  (2022) 063} [\href{https://arxiv.org/abs/2105.02862}{{\ttfamily
  2105.02862}}].

\bibitem{Rattazzi2008-je}
R.~Rattazzi, V.S.~Rychkov, E.~Tonni and A.~Vichi, \emph{Bounding scalar
  operator dimensions in {4D} {CFT}}, {\emph{J. High Energy Phys.} {\bfseries
  2008} (2008) 031}.

\bibitem{Bissi:2022mrs}
A.~Bissi, A.~Sinha and X.~Zhou, \emph{{Selected topics in analytic conformal
  bootstrap: A guided journey}},
  \href{https://doi.org/10.1016/j.physrep.2022.09.004}{\emph{Phys. Rept.}
  {\bfseries 991} (2022) 1} [\href{https://arxiv.org/abs/2202.08475}{{\ttfamily
  2202.08475}}].

\bibitem{Poland:2022qrs}
D.~Poland and D.~Simmons-Duffin, \emph{{Snowmass White Paper: The Numerical
  Conformal Bootstrap}},  in \emph{{Snowmass 2021}}, 3, 2022
  [\href{https://arxiv.org/abs/2203.08117}{{\ttfamily 2203.08117}}].

\bibitem{Hartman:2022zik}
T.~Hartman, D.~Mazac, D.~Simmons-Duffin and A.~Zhiboedov, \emph{{Snowmass White
  Paper: The Analytic Conformal Bootstrap}},  in \emph{{Snowmass 2021}}, 2,
  2022 [\href{https://arxiv.org/abs/2202.11012}{{\ttfamily 2202.11012}}].

\bibitem{Arkani-Hamed2018-dj}
N.~Arkani-Hamed, Y.-T.~Huang and S.-H.~Shao, \emph{On the positive geometry of
  conformal field theory},  \href{https://arxiv.org/abs/1812.07739}{{\ttfamily
  1812.07739}}.

\bibitem{Huang2019-yq}
Y.-T.~Huang, W.~Li and G.-L.~Lin, \emph{The geometry of optimal functionals},
  \href{https://arxiv.org/abs/1912.01273}{{\ttfamily 1912.01273}}.

\bibitem{Hartman:2014oaa}
T.~Hartman, C.A.~Keller and B.~Stoica, \emph{{Universal Spectrum of 2d
  Conformal Field Theory in the Large c Limit}},
  \href{https://doi.org/10.1007/JHEP09(2014)118}{\emph{JHEP} {\bfseries 09}
  (2014) 118} [\href{https://arxiv.org/abs/1405.5137}{{\ttfamily 1405.5137}}].

\bibitem{Benjamin:2016fhe}
N.~Benjamin, E.~Dyer, A.L.~Fitzpatrick and S.~Kachru, \emph{{Universal Bounds
  on Charged States in 2d CFT and 3d Gravity}},
  \href{https://doi.org/10.1007/JHEP08(2016)041}{\emph{JHEP} {\bfseries 08}
  (2016) 041} [\href{https://arxiv.org/abs/1603.09745}{{\ttfamily
  1603.09745}}].

\bibitem{Cardy:2017qhl}
J.~Cardy, A.~Maloney and H.~Maxfield, \emph{{A new handle on three-point
  coefficients: OPE asymptotics from genus two modular invariance}},
  \href{https://doi.org/10.1007/JHEP10(2017)136}{\emph{JHEP} {\bfseries 10}
  (2017) 136} [\href{https://arxiv.org/abs/1705.05855}{{\ttfamily
  1705.05855}}].

\bibitem{Cho:2017fzo}
M.~Cho, S.~Collier and X.~Yin, \emph{{Genus Two Modular Bootstrap}},
  \href{https://doi.org/10.1007/JHEP04(2019)022}{\emph{JHEP} {\bfseries 04}
  (2019) 022} [\href{https://arxiv.org/abs/1705.05865}{{\ttfamily
  1705.05865}}].

\bibitem{Dyer:2017rul}
E.~Dyer, A.L.~Fitzpatrick and Y.~Xin, \emph{{Constraints on Flavored 2d CFT
  Partition Functions}},
  \href{https://doi.org/10.1007/JHEP02(2018)148}{\emph{JHEP} {\bfseries 02}
  (2018) 148} [\href{https://arxiv.org/abs/1709.01533}{{\ttfamily
  1709.01533}}].

\bibitem{Collier:2017shs}
S.~Collier, P.~Kravchuk, Y.-H.~Lin and X.~Yin, \emph{{Bootstrapping the
  Spectral Function: On the Uniqueness of Liouville and the Universality of
  BTZ}}, \href{https://doi.org/10.1007/JHEP09(2018)150}{\emph{JHEP} {\bfseries
  09} (2018) 150} [\href{https://arxiv.org/abs/1702.00423}{{\ttfamily
  1702.00423}}].

\bibitem{Anous:2018hjh}
T.~Anous, R.~Mahajan and E.~Shaghoulian, \emph{{Parity and the modular
  bootstrap}},
  \href{https://doi.org/10.21468/SciPostPhys.5.3.022}{\emph{SciPost Phys.}
  {\bfseries 5} (2018) 022} [\href{https://arxiv.org/abs/1803.04938}{{\ttfamily
  1803.04938}}].

\bibitem{Mukhametzhanov:2019pzy}
B.~Mukhametzhanov and A.~Zhiboedov, \emph{{Modular invariance, tauberian
  theorems and microcanonical entropy}},
  \href{https://doi.org/10.1007/JHEP10(2019)261}{\emph{JHEP} {\bfseries 10}
  (2019) 261} [\href{https://arxiv.org/abs/1904.06359}{{\ttfamily
  1904.06359}}].

\bibitem{Benjamin:2019stq}
N.~Benjamin, H.~Ooguri, S.-H.~Shao and Y.~Wang, \emph{{Light-cone modular
  bootstrap and pure gravity}},
  \href{https://doi.org/10.1103/PhysRevD.100.066029}{\emph{Phys. Rev. D}
  {\bfseries 100} (2019) 066029}
  [\href{https://arxiv.org/abs/1906.04184}{{\ttfamily 1906.04184}}].

\bibitem{Mukhametzhanov:2020swe}
B.~Mukhametzhanov and S.~Pal, \emph{{Beurling-Selberg Extremization and Modular
  Bootstrap at High Energies}},
  \href{https://doi.org/10.21468/SciPostPhys.8.6.088}{\emph{SciPost Phys.}
  {\bfseries 8} (2020) 088} [\href{https://arxiv.org/abs/2003.14316}{{\ttfamily
  2003.14316}}].

\bibitem{Pal:2020wwd}
S.~Pal and Z.~Sun, \emph{{High Energy Modular Bootstrap, Global Symmetries and
  Defects}}, \href{https://doi.org/10.1007/JHEP08(2020)064}{\emph{JHEP}
  {\bfseries 08} (2020) 064}
  [\href{https://arxiv.org/abs/2004.12557}{{\ttfamily 2004.12557}}].

\bibitem{Benjamin:2020zbs}
N.~Benjamin and Y.-H.~Lin, \emph{{Lessons from the Ramond sector}},
  \href{https://doi.org/10.21468/SciPostPhys.9.5.065}{\emph{SciPost Phys.}
  {\bfseries 9} (2020) 065} [\href{https://arxiv.org/abs/2005.02394}{{\ttfamily
  2005.02394}}].

\bibitem{Dymarsky:2020bps}
A.~Dymarsky and A.~Shapere, \emph{{Solutions of modular bootstrap constraints
  from quantum codes}},
  \href{https://doi.org/10.1103/PhysRevLett.126.161602}{\emph{Phys. Rev. Lett.}
  {\bfseries 126} (2021) 161602}
  [\href{https://arxiv.org/abs/2009.01236}{{\ttfamily 2009.01236}}].

\bibitem{Lin:2021udi}
Y.-H.~Lin and S.-H.~Shao, \emph{{$\mathbb{Z}_N$ symmetries, anomalies, and the
  modular bootstrap}},
  \href{https://doi.org/10.1103/PhysRevD.103.125001}{\emph{Phys. Rev. D}
  {\bfseries 103} (2021) 125001}
  [\href{https://arxiv.org/abs/2101.08343}{{\ttfamily 2101.08343}}].

\bibitem{Benjamin:2021ygh}
N.~Benjamin, S.~Collier, A.L.~Fitzpatrick, A.~Maloney and E.~Perlmutter,
  \emph{{Harmonic analysis of 2d CFT partition functions}},
  \href{https://doi.org/10.1007/JHEP09(2021)174}{\emph{JHEP} {\bfseries 09}
  (2021) 174} [\href{https://arxiv.org/abs/2107.10744}{{\ttfamily
  2107.10744}}].

\bibitem{Grigoletto:2021zyv}
A.~Grigoletto and P.~Putrov, \emph{{Spin-Cobordisms, Surgeries and Fermionic
  Modular Bootstrap}},
  \href{https://doi.org/10.1007/s00220-023-04710-z}{\emph{Commun. Math. Phys.}
  {\bfseries 401} (2023) 3169}
  [\href{https://arxiv.org/abs/2106.16247}{{\ttfamily 2106.16247}}].

\bibitem{Benjamin:2022pnx}
N.~Benjamin and C.-H.~Chang, \emph{{Scalar modular bootstrap and zeros of the
  Riemann zeta function}},
  \href{https://doi.org/10.1007/JHEP11(2022)143}{\emph{JHEP} {\bfseries 11}
  (2022) 143} [\href{https://arxiv.org/abs/2208.02259}{{\ttfamily
  2208.02259}}].

\bibitem{Hellerman2011-su}
S.~Hellerman, \emph{A universal inequality for {CFT} and quantum gravity},
  {\emph{J. High Energy Phys.} {\bfseries 2011} (2011) 130}.

\bibitem{Friedan:2013cba}
D.~Friedan and C.A.~Keller, \emph{{Constraints on 2d CFT partition functions}},
  \href{https://doi.org/10.1007/JHEP10(2013)180}{\emph{JHEP} {\bfseries 10}
  (2013) 180} [\href{https://arxiv.org/abs/1307.6562}{{\ttfamily 1307.6562}}].

\bibitem{Collier:2016cls}
S.~Collier, Y.-H.~Lin and X.~Yin, \emph{{Modular Bootstrap Revisited}},
  \href{https://doi.org/10.1007/JHEP09(2018)061}{\emph{JHEP} {\bfseries 09}
  (2018) 061} [\href{https://arxiv.org/abs/1608.06241}{{\ttfamily
  1608.06241}}].

\bibitem{Hartman2019-dl}
T.~Hartman, D.~Maz{\'a}{\v c} and L.~Rastelli, \emph{Sphere packing and quantum
  gravity}, {\emph{J. High Energy Phys.} {\bfseries 2019} (2019) 48}.

\bibitem{Afkhami-Jeddi2019-ti}
N.~Afkhami-Jeddi, T.~Hartman and A.~Tajdini, \emph{Fast conformal bootstrap and
  constraints on 3d gravity}, {\emph{J. High Energy Phys.} {\bfseries 2019}
  (2019) 87}.

\bibitem{BTZ}
M.~Banados, C.~Teitelboim and J.~Zanelli, \emph{{The Black hole in
  three-dimensional space-time}},
  \href{https://doi.org/10.1103/PhysRevLett.69.1849}{\emph{Phys. Rev. Lett.}
  {\bfseries 69} (1992) 1849}
  [\href{https://arxiv.org/abs/hep-th/9204099}{{\ttfamily hep-th/9204099}}].

\bibitem{Kaidi:2020ecu}
J.~Kaidi and E.~Perlmutter, \emph{{Discreteness and integrality in Conformal
  Field Theory}}, \href{https://doi.org/10.1007/JHEP02(2021)064}{\emph{JHEP}
  {\bfseries 02} (2021) 064}
  [\href{https://arxiv.org/abs/2008.02190}{{\ttfamily 2008.02190}}].

\bibitem{Moment}
K.~Schm{\"u}dgen, \emph{The Moment Problem}, Springer, Cham (2017).

\bibitem{Chiang:2022ltp}
L.-Y.~Chiang, Y.-t.~Huang, L.~Rodina and H.-C.~Weng, \emph{{De-projecting the
  EFThedron}},  \href{https://arxiv.org/abs/2204.07140}{{\ttfamily
  2204.07140}}.

\bibitem{Fitzpatrick:2023lvh}
A.L.~Fitzpatrick and W.~Li, \emph{{Improving Modular Bootstrap Bounds with
  Integrality}},  \href{https://arxiv.org/abs/2308.08725}{{\ttfamily
  2308.08725}}.

\bibitem{Orbitopes}
R.~Sanyal, F.~Sottile and B.~Sturmfels, \emph{{Orbitopes}},
  \href{https://doi.org/10.1112/s002557931100132x}{\emph{Mathematika}
  {\bfseries 57} (2011) 275}.

\bibitem{Chiang:2022jep}
L.-Y.~Chiang, Y.-t.~Huang, W.~Li, L.~Rodina and H.-C.~Weng,
  \emph{{(Non)-projective bounds on gravitational EFT}},
  \href{https://arxiv.org/abs/2201.07177}{{\ttfamily 2201.07177}}.

\bibitem{Ginsparg:1987eb}
P.H.~Ginsparg, \emph{{Curiosities at c = 1}},
  \href{https://doi.org/10.1016/0550-3213(88)90249-0}{\emph{Nucl. Phys. B}
  {\bfseries 295} (1988) 153}.

\bibitem{DiUbaldo:2023hkc}
G.~Di~Ubaldo and E.~Perlmutter, \emph{{AdS$_3$ Pure Gravity and Stringy
  Unitarity}},  \href{https://arxiv.org/abs/2308.01787}{{\ttfamily
  2308.01787}}.

\bibitem{Keller:2012mr}
C.A.~Keller and H.~Ooguri, \emph{{Modular Constraints on Calabi-Yau
  Compactifications}},
  \href{https://doi.org/10.1007/s00220-013-1797-8}{\emph{Commun. Math. Phys.}
  {\bfseries 324} (2013) 107}
  [\href{https://arxiv.org/abs/1209.4649}{{\ttfamily 1209.4649}}].

\bibitem{Curto}
I.E.~Curto and L.A.~Fialkow, \emph{Recursiveness, positivity, and truncated
  moment problems}, {\emph{Houston J. Math} 603}.

\bibitem{nakata_2010}
M.~Nakata, \emph{A numerical evaluation of highly accurate multiple-precision
  arithmetic version of semidefinite programming solver: Sdpa-gmp, -qd and
  -dd.}, \href{https://doi.org/10.1109/cacsd.2010.5612693}{\emph{2010 IEEE
  International Symposium on Computer-Aided Control System Design} (2010) }.

\bibitem{simmons-duffin_2015}
D.~Simmons-Duffin, \emph{A semidefinite program solver for the conformal
  bootstrap}, \href{https://doi.org/10.1007/jhep06(2015)174}{\emph{Journal of
  High Energy Physics} {\bfseries 2015} (2015) }.

\bibitem{fishbonechiang}
L.-Y.~Chiang, ``Fishbonechiang/sdpjsolver.jl: A parallelized, arbitrary
  precision semidefinite program solver based on the primal-dual interior-point
  method..''

\end{thebibliography}\endgroup
\bibliographystyle{JHEP}

\end{document}